\documentclass{article}

\pdfoutput=1

\usepackage[left=2cm, right=2cm, top=2cm]{geometry}
\usepackage{amsmath}
\usepackage{amsfonts}
\usepackage{tabularx,multirow}
\usepackage{jcappub}
\usepackage{subcaption}
\usepackage[retainorgcmds]{IEEEtrantools}	
\usepackage{hyperref} 

\newcommand\cir[1]{\overset{o}{#1}}

\newcommand\Hbar{\bar{H}}
\newcommand\fchi{f^{\chi}}
\newcommand\fchip{f^{\chi\prime}}
\newcommand\fq{f^{q}}
\newcommand\fPi{f^{\Pi}}
\newcommand\feom{f^{eom}}
\newcommand\Hcal{\mathcal{H}}
\newcommand\Zcal{\mathcal{Z}}
\newcommand\gabar{\bar{\gamma}}
\newcommand\mn{\mu\nu}
\newcommand\Lcal{\mathcal{L}}
\newcommand\CAMB{\texttt{CAMB}}
\newcommand\CosmoMC{\texttt{CosmoMC}}
\newcommand\Hbarcal{\bar{\mathcal{H}}}

\title{\textbf{Observational status of the Galileon model general solution from cosmological data and gravitational waves}}
\author[a]{C. Leloup}
\author[a]{V. Ruhlmann-Kleider}
\author[a,b]{J. Neveu}
\author[a]{A. de Mattia}

\affiliation[a]{IRFU, CEA, Universit\'e Paris-Saclay, F-91191 Gif-sur-Yvette, France}
\affiliation[b]{LAL, Univ. Paris-Sud, CNRS/IN2P3, Universit\'e Paris-Saclay, Orsay, France}

\emailAdd{clement.leloup@cea.fr}
\emailAdd{vanina.ruhlmann-kleider@cea.fr}
\emailAdd{jneveu@lal.in2p3.fr}
\emailAdd{arnaud.de-mattia@cea.fr}

\abstract{The Galileon model is a tensor-scalar theory of gravity which explains the late acceleration of the Universe expansion with no instabilities and recovers General Relativity in the strong field limit. Most constraints obtained so far on Galileon model parameters from cosmological data were derived for the limited subset of tracker solutions and reported tensions between the model and data. This paper presents the first exploration of the general solution of the Galileon model, which is confronted against recent cosmological data for both background observables and linear perturbations, using Monte-Carlo Markov chains. As representative scenarios of the Galileon models, we study the Full galileon model with disformal coupling to matter and the uncoupled Cubic galileon model. We find that the general solution of the Full galileon model provides a good fit to CMB spectra, while the Cubic galileon model does not. When extending the comparison to BAO and SNIa data, even the general solution of the Full galileon model fails at providing a good fit to all datasets simultaneously. Tensions remain if the models are extended with an additional free parameter, such as the sum of active neutrino masses or the normalization of the CMB lensing spectrum. Finally, the multi-messenger observation of GW170817 is also discussed in the framework of the scenarios considered. The time delay between the gravitational signal and its electromagnetic counterpart was computed \textit{a posteriori} in every scenario of the Full galileon model cosmological fit chains and found to be ruled out by this observation.}

\keywords{Modified Gravity, Cosmological parameters from CMBR, Dark Energy theory}

\begin{document}
\maketitle
\flushbottom

\section{Introduction}
\label{section:section1}

Modifying standard General Relativity  (GR) by adding an additional scalar degree of freedom provides a physically motivated alternative to the cosmological constant to account for the late acceleration of the Universe expansion. Among modified gravity theories, the Galileon model offers a viable theoretical framework which is free from ghost degrees of freedom and fulfills the observational tests of gravity in the solar system. In this model, GR is supplemented by an additional scalar field, hereafter noted $\varphi$, whose equation of motion must be of second order in the field derivatives and invariant under the Galilean shift symmetry $\varphi \rightarrow  \varphi + c + b_\mu x^\mu$, where $c$ is a constant and $b_\mu$ is a constant vector.
The five possible Lagrangian terms for the uncoupled Galileon field $\varphi$ were derived in~\cite{Nicolis-2009} and were formulated in a covariant form in~\cite{Deffayet-2009a,Deffayet-2009b}. \\
\newline
The Galileon model has been extensively tested against various sets of cosmological data, whether in its uncoupled version (see e.g.~\cite{Appleby-2011,Barreira-2013,Neveu-2013}) or with direct couplings to matter (see~\cite{Neveu-2014,Neveu-2016}). Almost all recent results were restricted to the subset of tracker solutions of the model (see e.g~\cite{Barreira-2014,Barreira-2014b,Renk-2017,Peirone-2017}) after~\cite{Barreira-2013} advocated that tracker solutions were good representative of solutions able to reproduce the temperature CMB spectrum.
More recently, the observation of a gravitational wave (GW) event with an electromagnetic counterpart was used to reject Galileon models with quartic or quintic Lagrangian terms~\cite{Ezquiaga-2017,Wang-2017,Sakstein-2017}. \\
\newline
In this paper, we revisit the comparison of the Galileon model with cosmological data and test the general solution of the model, in order to assess the robustness of constraints derived from tracker solutions only. Best-fit scenarios are then confronted to GW data. \\
\newline
The outline of the paper is as follows. The main equations of the Galileon model are recalled in section~\ref{section:section2}. Section~\ref{section:section3} describes the datasets used to constrain the model while section~\ref{section:section4} introduces the methodology. Results are detailed in section~\ref{section:section5} and further discussed in section~\ref{section:section6}. We conclude in section~\ref{section:section7}.

\section{The Galileon model}
\label{section:section2}

The covariant action for the Galileon model with disformal coupling to matter is the following in the Jordan frame \cite{Nicolis-2009,Brax-2012}:
\begin{equation}
	\mathcal{S} = \int d^{4}x \sqrt{-g} \left[ \frac{M_{P}^{2}}{2} R - \frac{1}{2} \sum_{i=1}^{5} \frac{c_{i}}{M^{3 \left( i-2 \right)}} \mathcal{L}_{i} - \frac{M_{P}}{M^{3}} c_{G}G^{\mu\nu} \varphi_{;\mu} \varphi_{;\nu} - \mathcal{L}_{m} \right]
	\label{eq:action}
\end{equation}
where the $c_{i}$'s are dimensionless parameters, $R$ is the Ricci scalar, $M^{3} \equiv H_{0}^{2}M_{P}$ is a scale parameter, $G^{\mn}$ is the Einstein tensor, $\Lcal_{m}$ is the matter Lagrangian and the $\Lcal_{i}$'s are the Galileon Lagrangians defined by:
\begin{eqnarray}
\mathcal{L}_{1} & = & \varphi \\
\mathcal{L}_{2} & = & \varphi_{;\mu}\varphi^{;\mu} \equiv X \\
\mathcal{L}_{3} & = & X \Box \varphi \\
\mathcal{L}_{4} & = & X \left[ 2 \left( \Box \varphi \right)^{2} - 2 \left( \varphi_{;\mu\nu}\varphi^{;\mu\nu} \right) - \frac{1}{2}XR \right] \\
\mathcal{L}_{5} & = & X \left[ \left( \Box \varphi \right)^{3} - 3 \left( \varphi_{;\mu\nu}\varphi^{;\mu\nu} \right) \Box \varphi + 2 \left( \varphi_{;\mu}^{;\nu}\varphi_{;\nu}^{;\rho}\varphi_{;\rho}^{;\mu} \right) - 6 \left( \varphi_{;\mu}\varphi^{;\mu\nu}G_{\nu\rho}\varphi^{;\rho} \right) \right]
\end{eqnarray}
where $\Box = \nabla^{\mu}\nabla_{\mu}$ is the d'Alembertian operator. The semicolon expresses a covariant derivative with respect to space-time coordinates, also noted $\nabla_{\mu}$. $\Lcal_{1}$ is a tadpole term that acts as the usual cosmological constant, and may furthermore lead to vacuum instability because it describes an unbound potential term. Therefore, we set $c_{1} = 0$ in the following. $\Lcal_{2}$ is the usual kinetic term for a scalar field, while $\Lcal_{3}$ to $\Lcal_{5}$ are couplings of the Galileon field to itself and to the metric, bringing modifications to gravity. The Galileon model passes all solar system tests of gravity \cite{Burrage-2010} thanks to the Vainshtein screening effect 
introduced by the non-linear Lagrangians $\Lcal_{3}$ to $\Lcal_{5}$. \\
\newline
The direct disformal coupling between the Galileon field and matter appears in the Jordan frame as a coupling to the metric through the term $\propto G^{\mu\nu} \varphi_{;\mu} \varphi_{;\nu}$ with a coupling constant $c_{G}$. In principle, one could add a conformal coupling term $\propto \varphi R$ with a coupling constant $c_{0}$. However, it was shown in \cite{Neveu-2016} that scenarios with conformal coupling are disfavoured by cosmological distance observations. Furthermore, adding a conformal term to the action makes the numerical resolution of cosmological background and perturbations evolution much more complicated. Thus, we do not consider such a coupling in the following.

\subsection{Background evolution}
\label{subsection:subsection2.1}

Applying the variational principle to \eqref{eq:action} for the particular case of the FLRW metric, \cite{Appleby-2011} showed that the equations of motion for the cosmological background can be reduced to the following system of two coupled differential equations:
\begin{eqnarray}
	\cir{x} & = & -x + \frac{\alpha \lambda - \sigma \gamma}{\sigma \beta - \alpha \omega} \label{eq:ode bg 1} \\
	\cir{\Hbar} & = & \frac{\omega \gamma - \lambda \beta}{\sigma \beta - \alpha \omega}
	\label{eq:ode bg 2}
\end{eqnarray}
where the circle superscript corresponds to a derivative with respect to $\mathrm{ln} \left( a \right)$, $a$ being the scale factor of the FLRW metric. We defined the reduced Hubble rate $\Hbar \equiv H/H_{0}$, $x \equiv \cir{\varphi}/M_{P}$ and introduced the following functions:
\begin{eqnarray}
	\alpha & = & \frac{c_{2}}{6} \bar{H} x - 3 c_{3} \bar{H}^{3}x^{2}
+ 15c_{4} \bar{H}^{5}x^{3} - \frac{35}{2}c_{5} \bar{H}^{7}x^{4}
- 3c_{G} \bar{H}^{3}x \\
	\gamma & = & \frac{c_{2}}{3} \bar{H}^{2}x - c_{3} \bar{H}^{4}x^{2}
+ 5\frac{5}{2}c_{5} \bar{H}^{8}x^{4} - 2c_{G} \bar{H}^{4}x \\
	\beta & = & \frac{c_{2}}{6} \bar{H}^{2} - 2c_{3} \bar{H}^{4}x
+ 9c_{4} \bar{H}^{6}x^{2} - 10c_{5} \bar{H}^{8}x^{3}
- c_{G} \bar{H}^{4} \\
	\sigma & = & 2\bar{H} + 2c_{3} \bar{H}^{3}x^{3}
- 15c_{4} \bar{H}^{5}x^{4} + 21c_{5} \bar{H}^{7}x^{5}
+ 6c_{G} \bar{H}^{3}x^{2} \\
	\lambda & = & 3\bar{H}^{2} + \frac{\Omega_{\gamma}^{0}}{a^{4}}
+ \frac{p_{\nu}}{M_{Pl}^{2}H_{0}^{2}}
+ \frac{c_{2}}{2} \bar{H}^{2} x^{2} - 2c_{3} \bar{H}^{4}x^{3}
+ \frac{15}{2}c_{4} \bar{H}^{6}x^{4} - 9c_{5} \bar{H}^{8}x^{5}
- c_{G} \bar{H}^{4}x^{2} \\
	\omega & = & 2c_{3} \bar{H}^{4}x^{2}
- 12c_{4} \bar{H}^{6}x^{3} + 15c_{5} \bar{H}^{8}x^{4}
+ 4c_{G} \bar{H}^{4}x
\end{eqnarray}
Here, $\Omega_{\gamma}^{0}$ is the reduced energy density of the photons at present time, and $p_{\nu}$ is the neutrino pressure at any time which simply reduces to $\rho_{\nu}/3$ for massless neutrinos. \\
\newline
In order to compute the background evolution in the Galileon model, one needs to set the $c_{i}$'s and $c_{G}$ values, and to choose initial conditions for $\Hbar$ and $x$ at a given $z_{i}$. A convenient choice, following \cite{Neveu-2013}, is to take the initial condition at $z_{i} = 0$ for which $\Hbar \left( z_{i} = 0 \right) = 1$. The somehow difficult choice for $x_{0}$ will be treated in section \ref{section:section4}.

\subsection{Perturbation equations in the synchronous gauge}
\label{subsection:subsection2.2}

The covariant cosmological perturbation equations in Galileon cosmology were obtained in \cite{Barreira-2012} using a 3+1 decomposition. The idea of the 3+1 decomposition is to split physical quantities into their space and time components with regard to the 4-velocity $u^{\mu}$ of an observer to be defined when fixing the gauge. Space-like tensors can be obtained by projection on the 3-dimensional surfaces perpendicular to $u^{\mu}$ using the induced metric $\perp_{\mn} \equiv g_{\mn} + u_{\mu}u_{\nu}$. This projection operation will be denoted with a hat. For instance, the covariant spatial derivative of a tensor is given by:
\begin{equation}
    \hat{\nabla}^{\alpha}T^{\beta ... \gamma}_{\sigma ... \lambda} = \perp^{\alpha}_{\mu}\perp^{\beta}_{\nu} ... \perp^{\gamma}_{\kappa}\perp^{\rho}_{\sigma} ... \perp^{\eta}_{\lambda} \nabla^{\mu}T^{\nu ... \kappa}_{\rho ... \eta}
\end{equation}
It is then possible to express the energy-momentum tensor and the covariant derivative of the observer 4-velocity as:
\begin{eqnarray}
    T_{\mn} & = & \pi_{\mn} + 2q_{\left( \mu \right.}u_{\left. \nu \right)} + \rho u_{\mu}u_{\nu} + p\perp_{\mn} \\
    \nabla_{\mu}u_{\nu} & = & \sigma_{\mn} + \omega_{\mn} - u_{\mu}A_{\nu} + \frac{1}{3}\theta \perp_{\mn}
\end{eqnarray}
where $\pi_{\mn}$ is the Projected Symmetric and Trace-Free (PSTF) anisotropic stress tensor, $q_{\mu}$ is the heat flux vector, $\rho$ is the energy density, $p$ is the isotropic pressure, $\sigma_{\mn}$ is the PSTF shear tensor, $\omega_{\mn}$ the vorticity, $A_{\mu} \equiv u^{\mu}\nabla_{\mu}u_{\nu}$ is the acceleration of the observer and $\theta \equiv \nabla^{\alpha}u_{\alpha} = 3H$ is the expansion scalar. In order to obtain perturbed Einstein equations, one also needs to express the perturbations of the metric. We define the symmetric part $\mathcal{E}_{\mn}$ and anti-symmetric part $\mathcal{B}_{\mn}$ of the Weyl tensor $\mathcal{W}_{\mn\alpha\beta}$, which is the trace-free part of the Riemann tensor, and the projected Ricci scalar $\hat{R}$. \\
\newline
In the following, we only consider scalar modes of perturbations at linear order, since they are sufficient to compute the power spectra of the Cosmic Microwave Background. As shown in \cite{Challinor-1998}, scalar perturbations of $\omega_{\mn}$ and $\mathcal{B}_{\mn}$ are at most of second order and, thus, do not appear in scalar linear perturbation equations. Furthermore, the perturbation equations as implemented in the publicly available Einstein-Boltzmann solver \CAMB\ \cite{camb_notes} are written in the synchronous gauge \cite{Ma-1994} in which the acceleration $A_{\mu}$ vanishes. As our aim is to derive equations of Galileon perturbations suitable for CAMB, we adopt the same prescription in the following. We further decompose the remaining scalar perturbations in terms of Fourier space variables by the use of the eigenfunction $Q^{k}$ of the spatial Laplacian $\hat{\Box} \equiv \hat{\nabla}_{\mu}\hat{\nabla}^{\mu}$ that satisfies:
\begin{equation}
    \hat{\Box}Q^{k} = \frac{k^{2}}{a^{2}}Q^{k}
\end{equation}
Except for the pressure perturbations that are not used in \CAMB, the remaining scalar perturbations are decomposed on the $Q^{k}$ basis as:
\begin{IEEEeqnarray}{rClrCl}
    \hat{\nabla}_{\mu}\varphi & = & \sum_{k} M_{P}\frac{k}{a} \gamma Q_{\mu}^{k} \qquad & \qquad \hat{\nabla}_{\mu}\theta & = & \sum_{k}\frac{k^{2}}{a^{2}}\Zcal Q_{\mu}^{k} \\
	\hat{\nabla}_{\mu}\rho & = & \sum_{k}\frac{k}{a}\chi Q_{\mu}^{k} & q_{\mu} & = & \sum_{k}qQ_{\mu}^{k} \\
	\pi_{\mn} & = & \sum_{k}\Pi Q_{\mn}^{k} & \sigma_{\mn} & = & \sum_{k}\frac{k}{a}\sigma Q_{\mn}^{k} \\
	\hat{\nabla}_{\mu}\hat{R} & = & -4 \sum_{k}\frac{k^{3}}{a^{3}}\eta Q_{\mu}^{k} & \mathcal{E}_{\mn} & = & -\sum_{k}\frac{k^{2}}{a^{2}}\phi Q_{\mn}^{k}
\end{IEEEeqnarray}
The scalar perturbations extracted from the energy-momentum tensor are decomposed into their contributions from ordinary matter and from the Galileon field:
\begin{equation}
    \chi = \chi^{m} + \chi^{G}, \quad q = q^{m} + q^{G} \quad \text{and} \quad \Pi = \Pi^{m} + \Pi^{G}
\end{equation}
The scalar perturbation equations for the Galileon quantities and the equation of motion of the Galileon field scalar perturbation are given in \cite{Barreira-2012}. With our notations, they can be expressed as:
\begin{eqnarray}
	&& \chi^{G} = \fchi_{1} \cdot \gamma + \fchi_{2} \cdot \gamma'  + \frac{1}{\kappa a^{2}} \left( \fchi_{3} \cdot k\Hcal\Zcal + \fchi_{4} \cdot k^{2}\eta \right) \label{eq:chiG 1}\\
	&& q^{G} = \fq_{1} + \frac{1}{\kappa a^{2}} \fq_{2} \cdot k^{2} \left( \sigma - \Zcal \right) \\
	&& \Pi^{G} = \fPi_{1} + \frac{1}{\kappa a^{2}} \left( \fPi_{2} \cdot k\Hcal\sigma - \fPi_{3} \cdot k\sigma' + \fPi_{4} \cdot k^{2}\phi \right) \\
	&& 0 = \feom_{1} \cdot \gamma'' + \feom_{2} \cdot \gamma' + \feom_{3} \cdot k^{2}\gamma + \feom_{4} \cdot k\Hcal\Zcal + \feom_{5} \cdot k\Zcal' + \feom_{6} \cdot k^{2}\eta \label{eq:eom 1}
\end{eqnarray}
The expressions for the $f^{X}_{i}$ functions can be found in appendix \ref{appendix:appendixA}. We also introduced $\Hcal = \frac{a'}{a}$, where the prime corresponds to a derivative with respect to the conformal time $\tau$. However, the scalar perturbations that appear on the right hand sides of \eqref{eq:chiG 1} to \eqref{eq:eom 1} depend themselves on Galileon quantities through the Einstein equations in the synchronous gauge \cite{Ma-1994}. 
After a bit of algebra, one can obtain expressions that are suitable for a step by step numerical resolution of the evolution of scalar perturbations:
\begin{eqnarray}
	&& \chi^{G} = \frac{2}{2-\fchi_{3}} \left[ \fchi_{1} \cdot \gamma + \fchi_{2} \cdot \gamma' + \frac{\fchi_{3}}{2} + \frac{\fchi_{3} \cdot \chi^{f} - 2\fchi_{4}}{\kappa a^{2}} \cdot k^{2}\eta \right] \label{eq:chiG CAMB}\\
	&& q^{G} = \frac{2}{2-3\fq_{2}} \left[ \fq_{1} + \frac{3}{2}\fq_{2} \cdot q^{f} \right] \\
	&& \Pi^{G} = \frac{2}{2-2\fPi_{3}+\fPi_{4}} \left[ \fPi_{1} + \frac{2\fPi_{3}-\fPi_{4}}{2}\Pi^{f} \right. \nonumber \\
	&& \left. + \frac{\fPi_{2}+2\fPi_{3}-\fPi_{4}}{2} \left( \chi + \frac{3}{k}\Hcal q \right) + \frac{\fPi_{2}+2\fPi_{3}-\fPi_{4}}{2} \left( \chi + \frac{3}{k}\Hcal q \right)+ \left( \fPi_{2} + \fPi_{3} \right) k^{2}\eta \right] \\
	&& \gamma'' = -\frac{1}{\feom_{1} + \fchi_{2}\xi} \left[ \xi \left( \kappa a^{2}\chi^{f\prime} + \kappa a^{2}f^{\chi\prime}_{1}\gamma + \kappa a^{2}\fchi_{1}\gamma' + \kappa a^{2}f^{\chi\prime}_{2}\gamma' + \left( 1-\fchi_{4} \right) k \kappa a^{2}q \right) \right. \nonumber \\
	&& \left. + \feom_{3}\Hcal\gamma' + \feom_{3}k^{2}\gamma + \left( \frac{\feom_{4}}{2} + \xi \left[ \frac{\kappa a^{2} \left( \bar{\rho} + \bar{p} \right)}{2\Hcal} + \Hcal - \fchi_{3}\Hcal + \frac{\fchip_{2} + \fchi_{3}\cir{\Hcal}}{2} \right] \right) \kappa a^{2}\chi \right. \nonumber \\
	&& \left. + \left( \feom_{4} - 2\feom_{6} + \xi \left[ \frac{\kappa a^{2} \left( \bar{\rho} + \bar{p} \right)}{\Hcal} - 2\Hcal + \fchip_{2} + \fchi_{3}\cir{\Hcal} - 2\fchi_{3}\Hcal - 2\fchip_{3} + 4\fchi_{4}\Hcal \right] \right) k^{2}\eta \right] \label{eq:eom CAMB}
\end{eqnarray}
where we have defined $\xi = \frac{\feom_{5}}{\left( 2-\fchi_{3} \right) \Hcal}$. In order to solve the differential equations, one only need two additional initial conditions for $\gamma$ and $\gamma'$, taken at very early time, following the prescription for \CAMB \ initial conditions of the perturbations. However, it was noted in \cite{Barreira-2012} that the exact value for these initial conditions does not influence significantly the evolution of Galileon perturbations and can be taken equal to 0, which will be done in all the following results. 

\section{Datasets}
\label{section:section3}

\subsection{Type Ia Supernovae}
\label{subsection:subsection3.1}

In this work, we used the Joint Light-curve Analysis (JLA) sample of type Ia supernovae (SNIa) published jointly by the Supernova Legacy Survey (SNLS) and the Sloan Digital Sky Survey (SDSS) collaboration \cite{Betoule-2014}. The JLA catalog is composed of 740 SNIa, among which 374 were observed by the SDSS telescope, 242 come from the first three years of SNLS data observed by the Canada-France-Hawaii Telescope, 110 are low redshift events ($z < 0.08$) 
from several surveys
and finally 14 are high redshift SNIa ($0.7 < z < 1.4$) observed by the Hubble Space Telescope (HST). We used the published JLA results, including covariance matrices.

\subsection{Baryonic Acoustic Oscillations}
\label{subsection:subsection3.2}

The Baryonic Acoustic Oscillations (BAO) scale measurements used here come from several sources. At low redshift, we used a measurement at an average redshift of $z=0.106$ from the 6dF Galaxy Survey (6dFGS) catalog of galaxies \cite{Beutler-2011} and a measurement at an average redshift of $z=0.15$ from the SDSS DR7 Main Galaxy Sample (MGS) \cite{Ross-2014}. Finally, we used three additional measurements at average redshifts of $z=0.38,0.51,0.61$ from the galaxy sample of BOSS DR12 \cite{Alam-2016}. We used the published covariance matrix for the last three measurements as they are not independent. \\
\newline
The BOSS DR12 measurements probe the BAO scale at a given redshift in an anisotropic manner, which measures the Hubble rate and the comoving angular diameter distance defined by:
\begin{equation}
    D_{M} \left( z \right)  = \int_{0}^{z} \frac{cdz'}{H \left( z' \right)}
\end{equation}
Because of limited sample size, the low redshift measurements were performed assuming an isotropic behaviour, which can only constrain an effective distance parameter defined by:
\begin{equation}
    D_{V} \left( z \right)  = \left( D_{M}^{2} \left( z \right) \frac{cz}{H \left( z\right)} \right)^{1/3}
\end{equation}
BAO observables are sensitive to the above distances and Hubble rate normalized by $r_{d}$, the sound horizon at the drag epoch. We used the BAO measurements summarized in table \ref{table:BAO data}. They correspond to the baseline of the Planck Collaboration in 2015 \cite{Planck-2015} that we decided to follow.
\begin{table}[htbp]
\begin{center}
\scalebox{0.9}{
\begin{tabularx}{\textwidth}{|c|cccc|}
    \cline{1-5}
    \rule{0pt}{2.5ex} Data & $z_{eff}$ & $D_{V} \left( r_{d, fid}/r_{d} \right) \left[ \text{Mpc} \right]$ & $D_{M} \left( r_{d, fid}/r_{d} \right) \left[ \text{Mpc} \right]$ & $H \left( r_{d}/r_{d, fid} \right) \left[ \text{km s}^{-1} \text{Mpc}^{-1} \right]$ \\
    \cline{1-5}
    \rule{0pt}{3ex} 6dFGS \cite{Beutler-2011} & $0.106$ & $456 \pm 27$ & - & - \\
    	\cline{1-5}
    \rule{0pt}{3ex} MGS \cite{Ross-2014} & $0.15$ & $664 \pm 25$ & - & - \\
    	\cline{1-5}
    \rule{0pt}{3ex} \multirow{3}*{BOSS \ DR12 \cite{Alam-2016}}
    & $0.38$ & - & $1512 \pm 33$ & $81.2 \pm 3.2$ \\ 
    & $0.51$ & - & $1975 \pm 41$ & $90.9 \pm 3.2$ \\ 
    & $0.61$ & - & $2307 \pm 50$ & $99.0 \pm 3.4$ \\
    \cline{1-5}
\end{tabularx}
}
\caption{Summary of BAO measurements used in this work. They come from several sources and are given as measurements of $D_{V}$ (isotropic analyses) or of both $D_{M}$ and $H$ (anisotropic analyses), normalized by the ratio of $r_{d}$, the sound horizon at the drag epoch, to its value in the fiducial cosmology used in the analysis, $r_{d, fid}$.}
\label{table:BAO data}
\end{center}
\end{table}

\subsection{Cosmic Microwave Background}
\label{subsection:subsection3.3}

The Cosmic Microwave Background (CMB) data used in this paper are those published by the Planck Collaboration in 2015 \cite{Planck-2015} that will be referred to as Planck 2015 in the following. We used all available binned data from Planck 2015, namely the temperature power spectrum (TT), the polarization spectrum from E modes (EE), the cross-spectrum between temperature and polarization E modes (TE) and finally the lensing power spectrum. The first three power spectra extend up to $\ell \leq 2508$. Following the Planck collaboration prescription, we restrict the use of the lensing power spectrum to the multipole range where the reconstruction was proved to be robust and the estimator signal-to-noise to be greatest, namely $40 \leq \ell \leq 400$.

\subsection{Gravitational Waves}
\label{subsection:subsection3.4}

On the 17th of August 2017, a gravitational wave signal was observed with an electromagnetic counterpart from what was interpreted to be a neutron star binary merger \cite{GBM-2017}. This event, referred to as GW170817 in the following, puts strong constraints on modified gravity models since many of them predict GW and light to have different propagation speeds. This is, in particular, the case of the Galileon model, for which the GW speed is given by \cite{DeFelice-2011}:
\begin{equation}
    c_{g}^{2} = \frac{\frac{1}{2} + \frac{1}{4}c_{4}\Hbar^{4}x^{4} + \frac{3}{2}c_{5}\Hbar^{5}x^{4}\cir{\left( \Hbar x \right)} - \frac{1}{2}c_{G}\Hbar^{2}x^{2}}{\frac{1}{2} - \frac{3}{4}c_{4}\Hbar^{4}x^{4} + \frac{3}{2}c_{5}\Hbar^{5}x^{5} + \frac{1}{2}c_{G}\Hbar^{2}x^{2}}
\end{equation}
Note that $c_g$ differs from the speed of light only if at least one of the parameters $c_4, c_5, c_G$ is non zero. The useful observable is the time delay between the arrival of the GW and the corresponding Gamma Ray Burst, measured to be $\Delta t = 1.74 \pm 0.05 \ \mathrm{s}$ \cite{GBM-2017}. For a given model, assuming there is no time delay between GW and light emissions, the time delay at arrival $\Delta t$ is defined by:
\begin{equation}
    \Delta t = \frac{1}{H_{0}} \int_{a_{e}}^{1} \frac{da}{a\bar{H}} \left( 1 - \frac{c}{c_{g} \left( a \right)} \right)
\end{equation}
In principle, we should consider a nuisance parameter $\delta t$ to take into account the time delay at emission, assumed to be less than $10 \ \mathrm{s}$.

\section{Methodology}
\label{section:section4}


Predictions for the cosmological observables corresponding to the above datasets depend on a few free parameters. The first set of free parameters is common to most cosmological models:
\begin{equation}
    \lbrace \Omega_{b}h^{2}, \Omega_{c}h^{2}, 100 \theta_{MC}, \tau, n_{s}, A_{s} \rbrace
\end{equation}
Here, $\Omega_{b}$ is the density of baryonic matter at present time, $\Omega_{c}$ the density of cold dark matter at present time, $h$ the dimensionless Hubble constant defined by $H_{0}=h \cdot 100 \ \text{km} \cdot \text{s}^{-1} \cdot \text{Mpc}$, $\theta_{MC}$ the angular scale of the CMB first peak, $\tau$ the optical depth at reionization, $n_{s}$ the spectral index of the primordial fluctuations and $A_{s}$ the amplitude of these primordial fluctuations. These parameters are the only free parameters of the base $\Lambda$CDM model.
In addition to these cosmological free parameters, the Galileon model has its own parameters: 
\begin{equation}
    \lbrace c_{2}, c_{3}, c_{4}, c_{5}, c_{G}, x_{0} \rbrace
\end{equation}
However, it has been shown in \cite{DeFelice-2010} that the system of differential equations \eqref{eq:ode bg 1} and \eqref{eq:ode bg 2} for background evolution is invariant under the following transformation of the $c_{i}$'s and $x$, for any arbitrary parameter $A$:
\begin{eqnarray}
    c_{i} & \rightarrow & c_{i} \equiv c_{i} A ^{i}, \quad i=2,...,5 \\
    c_{G} & \rightarrow & c_{G} \equiv c_{G} A ^{2} \\
    x & \rightarrow & x \equiv x/A \\
    \gamma & \rightarrow & \gamma \equiv \gamma/A
\end{eqnarray}
Following~\cite{Neveu-2013}, we choose to set $A$ to the value of $x$ at present time $x_{0}$, to break the degeneracy between the Galileon parameters. This rescaling avoids to choose the initial value $x_{0}$ to solve the system of differential equations, since the rescaled function $x$ has the simple initial condition $x_{0}=1$, regardless of the scenario. Thus, the only free parameters for the Galileon model are:
\begin{equation}
    \lbrace c_{2}, c_{3}, c_{4}, c_{5}, c_{G} \rbrace
\end{equation}
It is possible to set one of these imposing the Universe to be flat at present time. Flatness translates into a new relation between the $c_{i}$'s that comes from the first Friedmann equation:
\begin{equation}
    1 = \Omega_{m}^{0} + \Omega_{\gamma}^{0} + \Omega_{\nu}^{0} + \frac{c_{2}}{6} - 2c_{3} + \frac{15}{2}c_{4} - 7c_{5} - 3c_{G}
\end{equation}
This equation will be used to set the value of the parameter that corresponds to the highest order terms in the background system of equations \eqref{eq:ode bg 1} and \eqref{eq:ode bg 2}. \\
\newline
We define three categories of models to constrain:
\begin{itemize}
    \item[] \textbf{$\Lambda$CDM}, as a reference, with the set of free parameters $\lbrace \Omega_{b}h^{2}, \Omega_{c}h^{2}, 100 \theta_{MC}, \tau, n_{s}, A_{s} \rbrace$
    \item[] \textbf{Full galileon}, which corresponds to the complete Galileon model described in section \ref{section:section2}. The set of free parameters for this model is $\lbrace \Omega_{b}h^{2}, \Omega_{c}h^{2}, 100 \theta_{MC}, \tau, n_{s}, A_{s}, c_{2}, c_{3}, c_{4}, c_{G} \rbrace$
    \item[] \textbf{Cubic galileon}, which is a subset of the previous model where $c_{4}=c_{5}=c_{G}=0$. The set of free parameters for this model is $\lbrace \Omega_{b}h^{2}, \Omega_{c}h^{2}, 100 \theta_{MC}, \tau, n_{s}, A_{s}, c_{2} \rbrace$. This subset of models is particularly interesting because as $c_{4}=c_{5}=c_{G}=0$, the GW speed is equal to the speed of light.
\end{itemize}
Following the terminology of the Planck Collaboration \cite{Planck-2015}, the above models will be referred to as base models. We will also study extended models for which one additional free parameter is considered, namely the normalization of the lensing power spectrum $A_{L}$ \cite{Planck-2015} or the sum of active neutrino masses $\sum m_{\nu}$. These two parameters are set to $A_{L} = 1$ and $\sum m_{\nu}=0.06 \ \mathrm{eV}$ in the base models. In both base and extended models, all other parameters are set to their \CAMB \ default values. \\
\newline
To derive constraints on the above models, we modified the version of \CAMB \ \cite{camb_notes} released in November 2016 to include the Galileon cosmology in addition to the standard $\Lambda$CDM cosmology. We have thoroughly checked that the results for the background and perturbation evolutions in Galileon cosmology from our version of \CAMB \ are compatible with independent previous results. In particular, the angular power spectrum of temperature anisotropies and the matter power spectrum were found to be compatible at better than 1\% with those published in \cite{Barreira-2012}. The exploration of the model parameter space was performed using our own modified version\footnote{Our modified versions of CosmoMC, that includes the modified version of CAMB and the Galileon free and derived parameters, is freely available at \url{https://github.com/ClementLeloup/cosmomc_galileon}.} of the November 2016 release of the publicly available Monte-Carlo Markov Chain (MCMC) code  \CosmoMC\ \cite{Lewis-2002}. More details on the parameter exploration can be found in appendix \ref{appendix:appendixB}.

\section{Results}
\label{section:section5}

\subsection{Base models}
\label{subsection:subsection5.1}

\begin{figure}[htbp]
\begin{center}
\includegraphics[width=\textwidth]{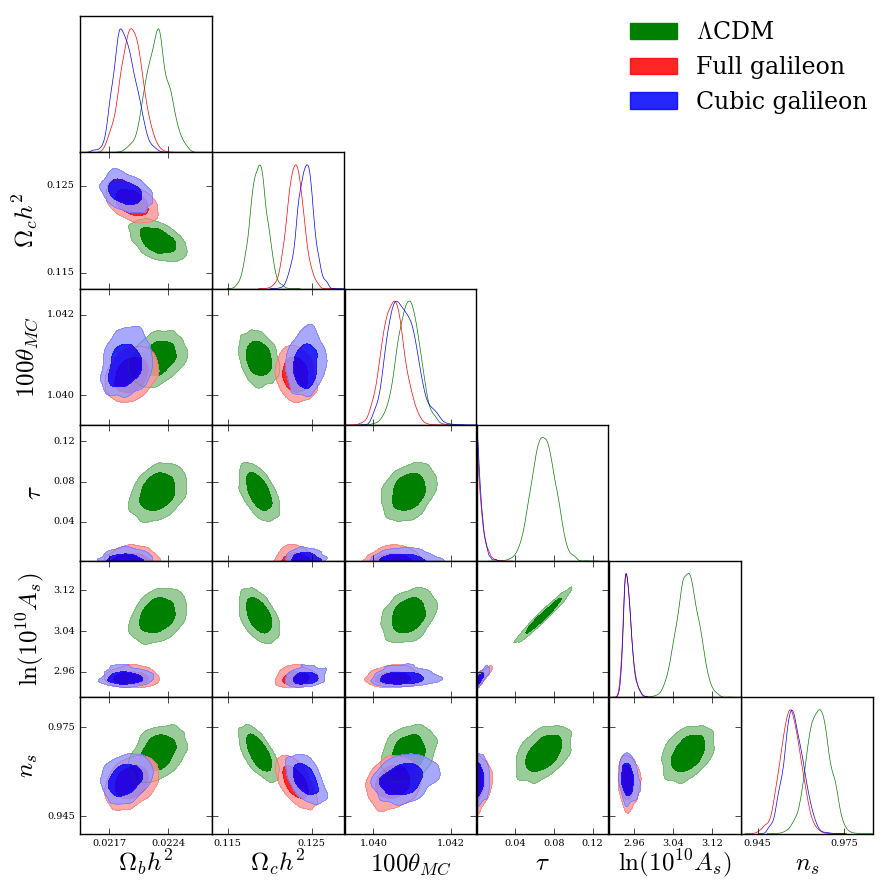}
\caption{Marginalized 1D and 2D constraints on the common parameters to the three base models from fits to CMB, BAO and JLA data combined.}
\label{fig:CMB+BAO+JLA usual}
\end{center}
\end{figure}
\begin{figure}[htbp]
\begin{center}
\begin{subfigure}{.5\linewidth}
  \centering
  \includegraphics[width=\linewidth]{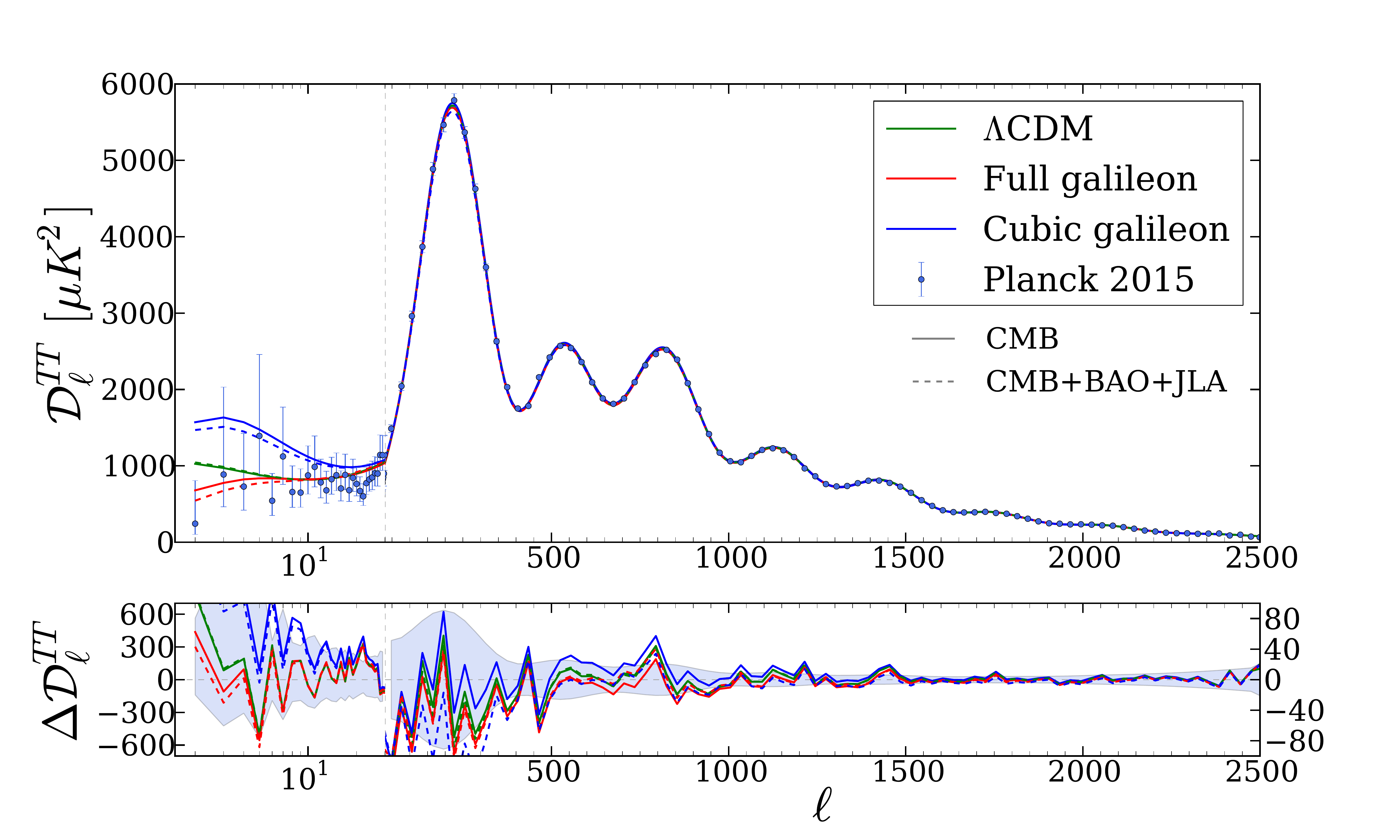}
\end{subfigure}%
\begin{subfigure}{.5\linewidth}
  \centering
  \includegraphics[width=\linewidth]{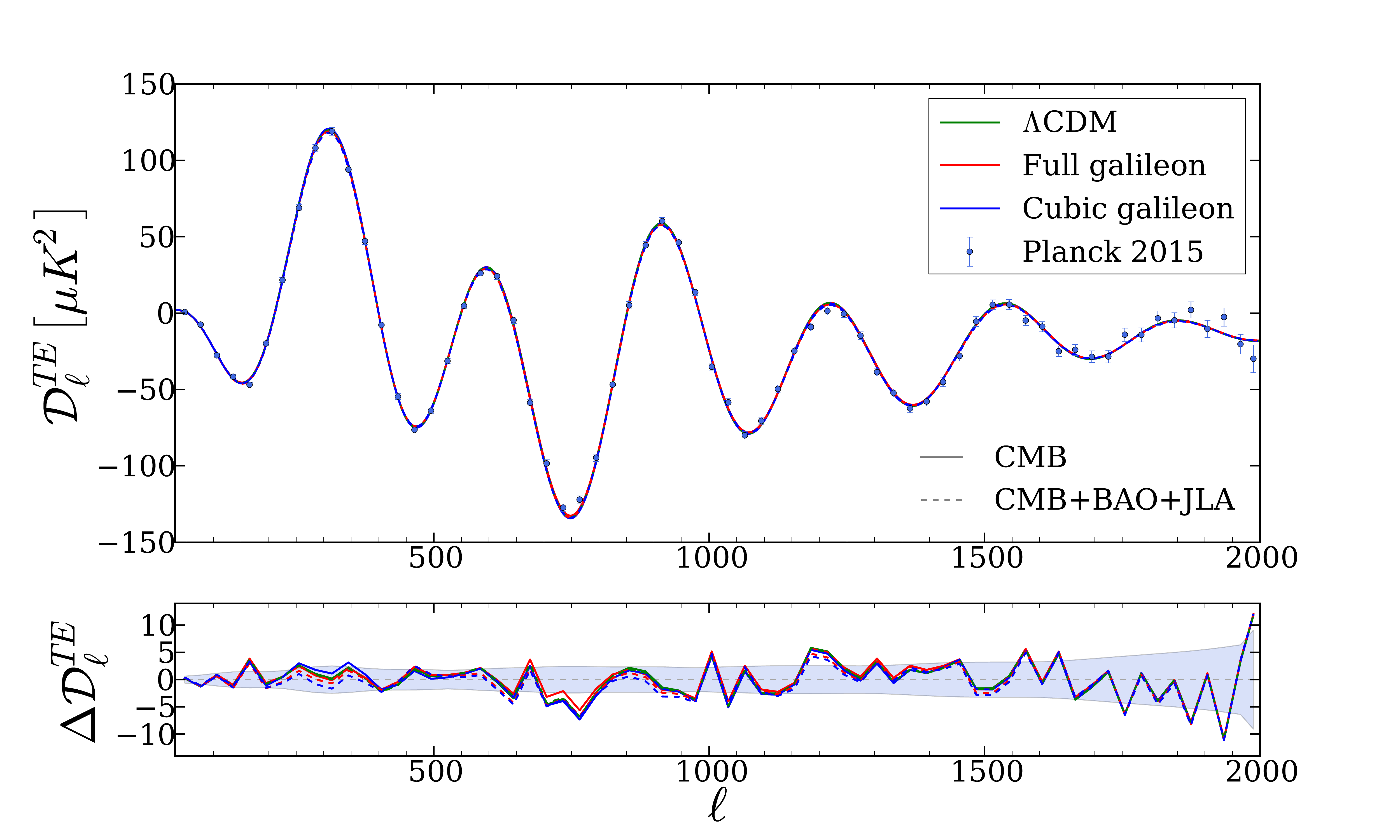}
\end{subfigure}
\begin{subfigure}{.495\linewidth}
  \centering
  \includegraphics[width=\linewidth]{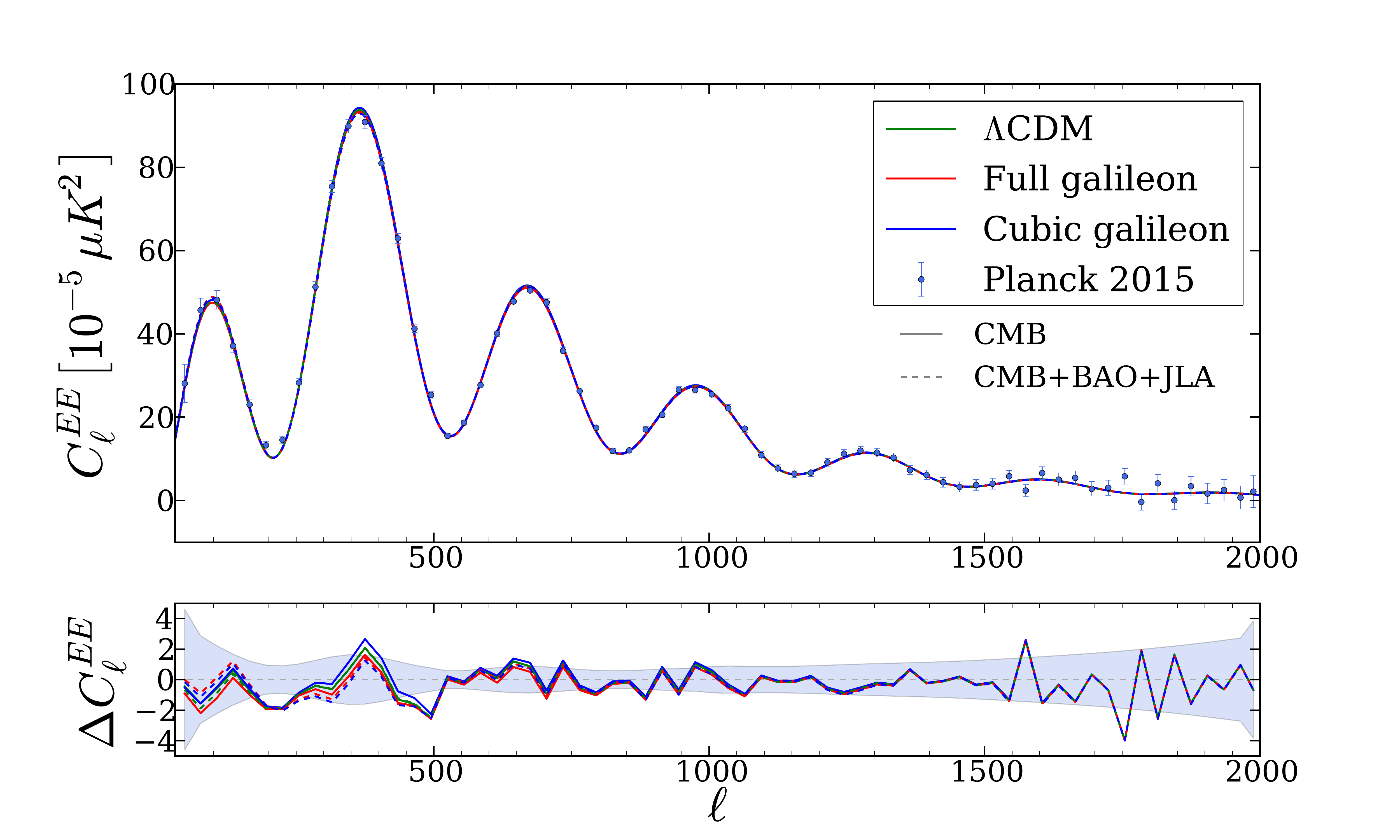}
\end{subfigure}
\begin{subfigure}{.495\linewidth}
  \centering
  \includegraphics[width=\linewidth]{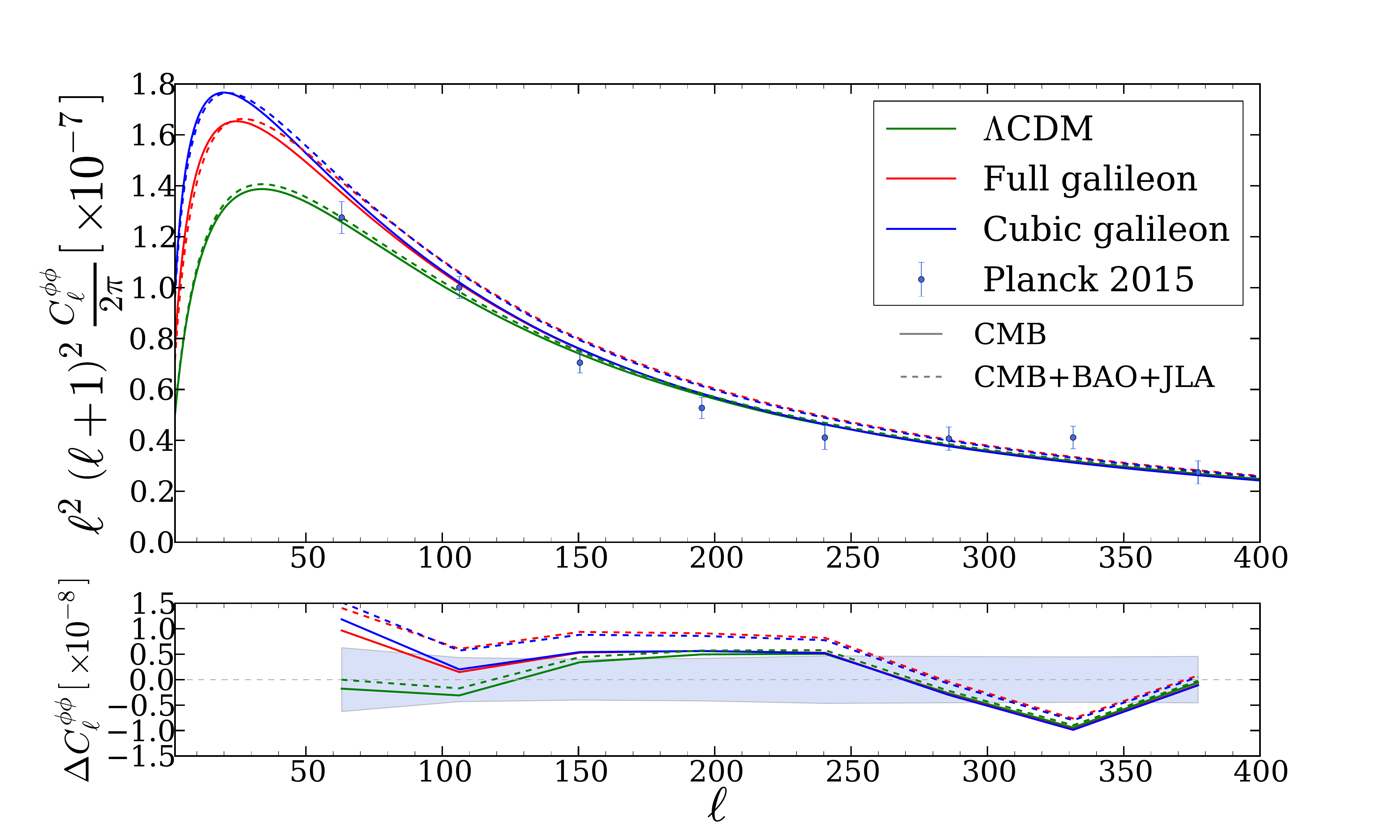}
\end{subfigure}
\end{center}
\caption{CMB power spectra of the best-fit to CMB data only (solid) and to the combination of CMB, BAO and JLA data (dashed) in the three base models, compared to Planck 2015 data (dots).}
\label{fig:CMB+BAO+JLA best-fit CMB}
\end{figure}
\begin{figure}[htbp]
\begin{center}
\includegraphics[width=\textwidth]{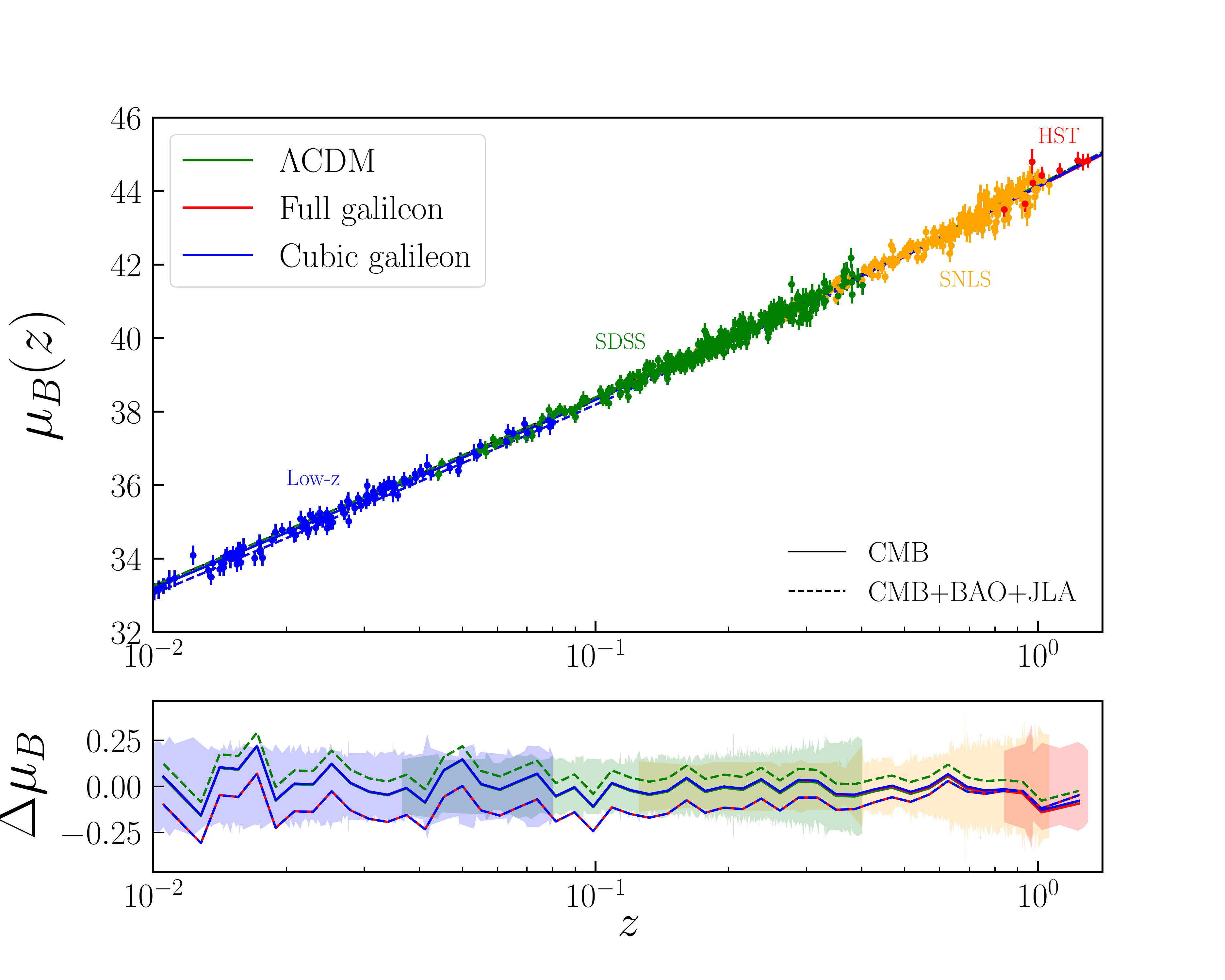}
\caption{(Top) Hubble diagram of the JLA SNIa sample with luminosity distance predictions of the best-fit to JLA data only (solid) and to the combination of CMB, BAO and JLA data (dashed) in the three base models. (Bottom) Difference between observations and 
predictions averaged on logarithmic bins in $z$ for the three base models. The colored areas represent the observational uncertainties.}
\label{fig:CMB+BAO+JLA bestfit hubble}
\end{center}
\end{figure}
\begin{figure}[htbp]
\begin{center}
\begin{subfigure}{0.5\textwidth}
  \centering
  \includegraphics[width=\linewidth]{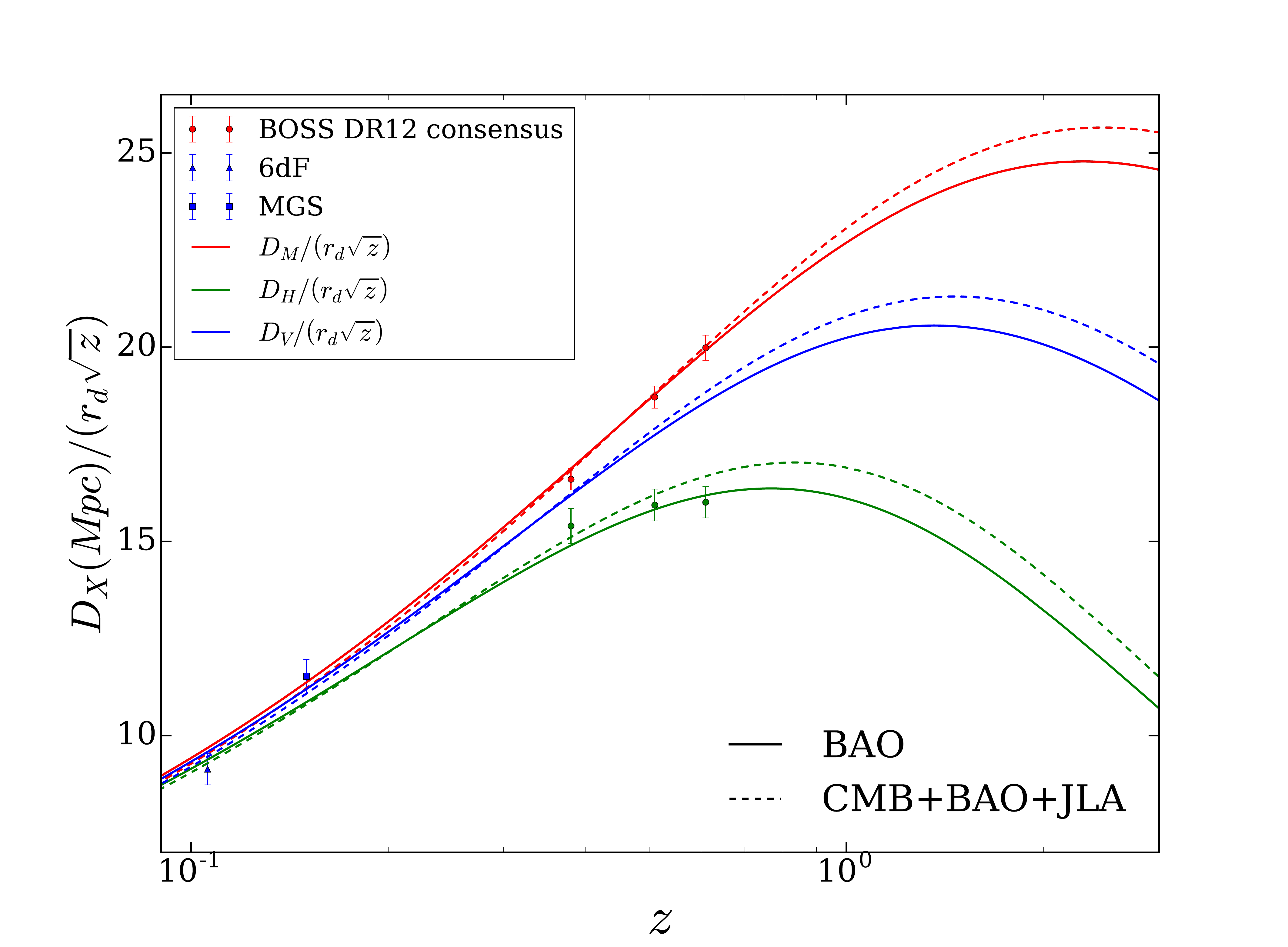}
\end{subfigure}%
\begin{subfigure}{0.5\textwidth}
  \centering
  \includegraphics[width=\linewidth]{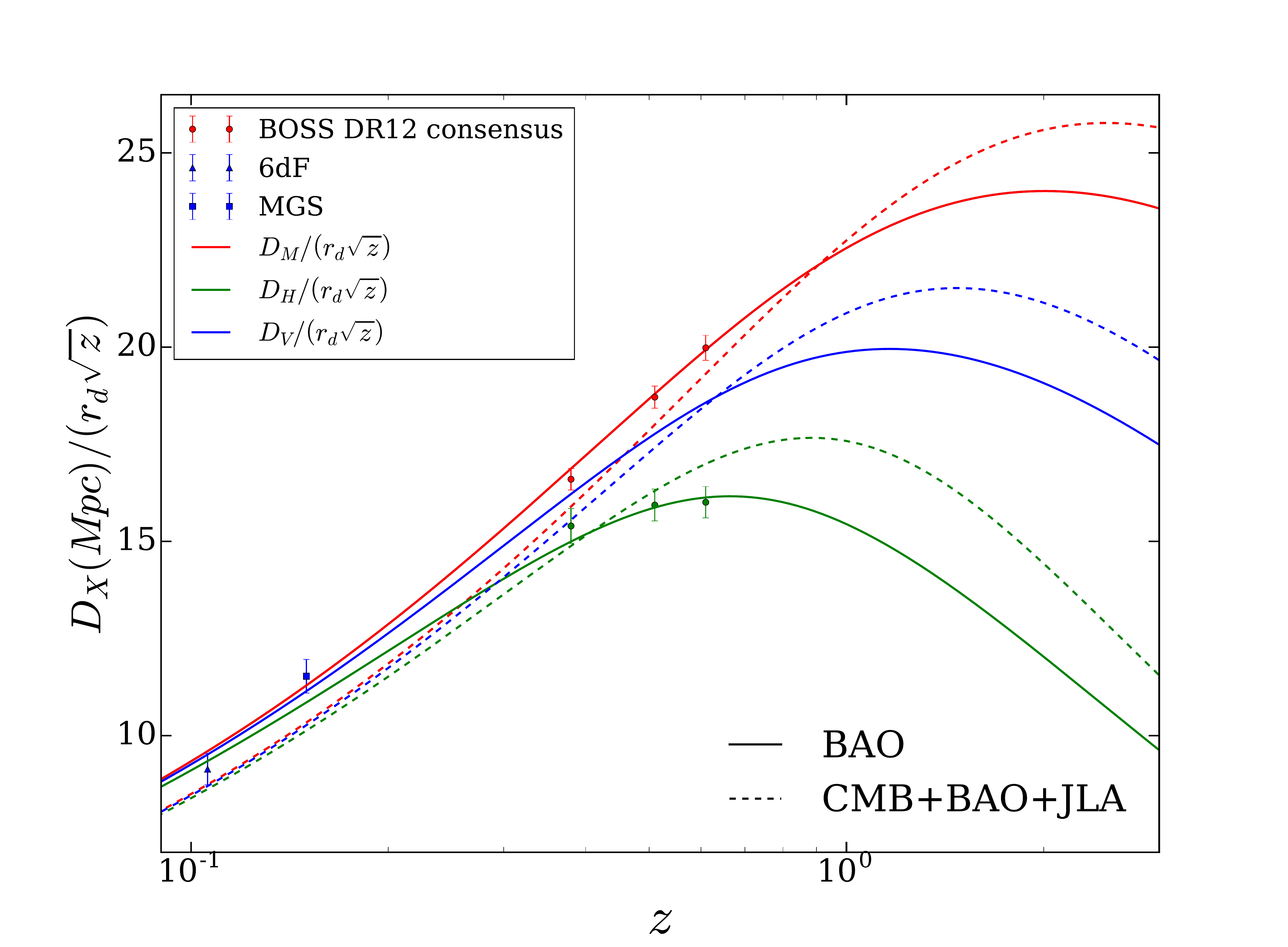}
\end{subfigure}
\begin{subfigure}{0.5\textwidth}
  \includegraphics[width=\linewidth]{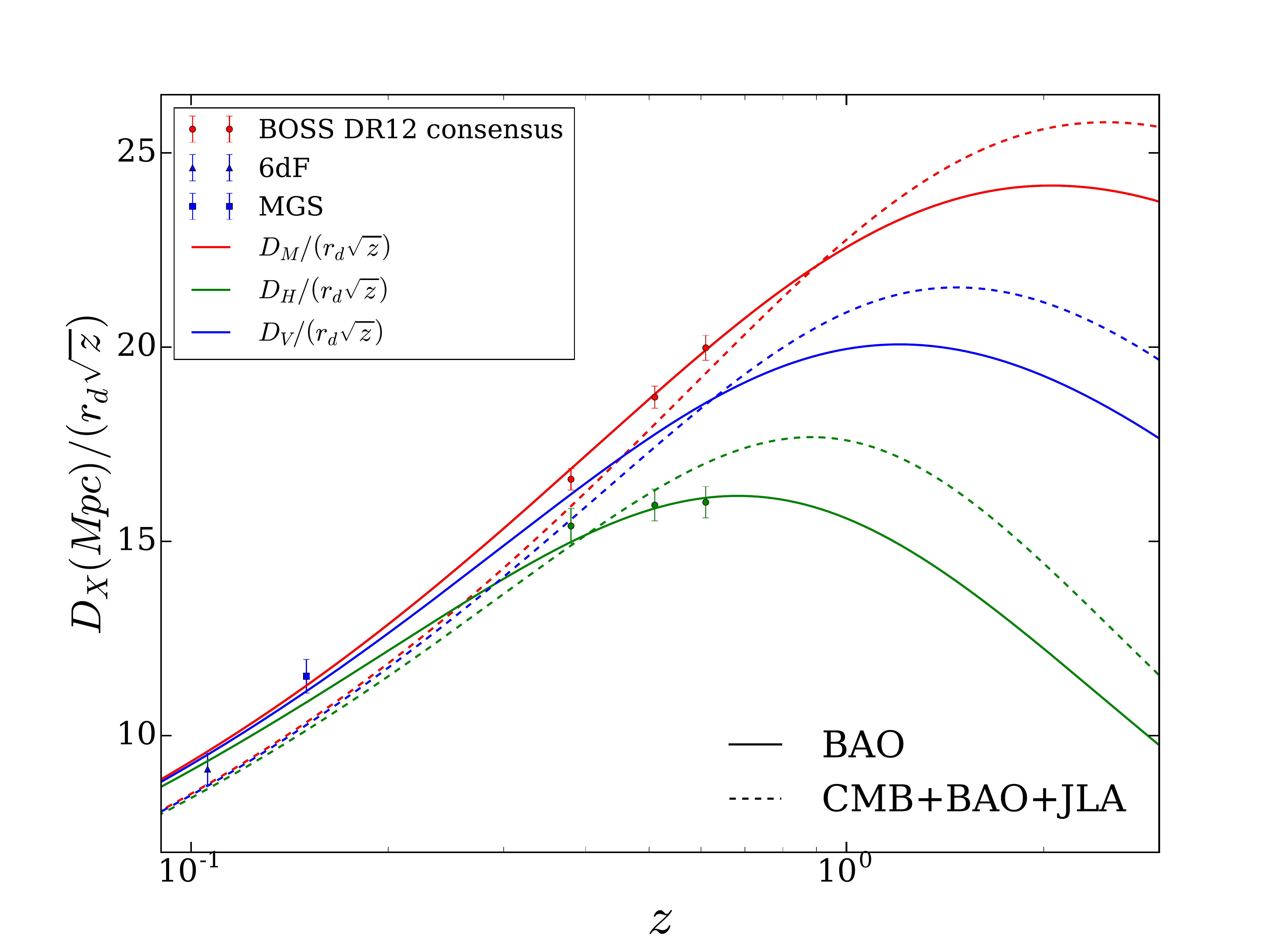}
\end{subfigure}
\end{center}
\caption{$D_{M}$, $D_{V}$ and $D_{H}$ predictions of the best-fit to BAO data only (solid) and to the combination of CMB, BAO and JLA data (dashed) for the $\Lambda$CDM model (top left), Full galileon model (top right) and Cubic galileon model (bottom).}
\label{fig:CMB+BAO+JLA best-fit BAO}
\end{figure}

Figure \ref{fig:CMB+BAO+JLA usual} shows the one-dimensional marginalized likelihood ($\Lcal$) and the two-dimensional $1\sigma$ and $2\sigma$ contours obtained using the combination of CMB, BAO and JLA data for the parameters common to the three base models. Table \ref{table:Contraintes CMB+BAO+JLA} summarizes the one-dimensional marginalized likelihood values of the parameters of the three models along with the values of $\chi^{2} = -2 \mathrm{ln} \Lcal$ for the set of parameters maximizing the likelihood $\Lcal$ of the combined fits. Figure \ref{fig:CMB+BAO+JLA best-fit CMB} shows the TT, TE, EE and lensing power spectra for the best fitting parameters to CMB data only and to the combination of CMB, BAO and JLA data, for the three models. 
Figure \ref{fig:CMB+BAO+JLA bestfit hubble} and Figure \ref{fig:CMB+BAO+JLA best-fit BAO} are the equivalent figures for the SNIa Hubble diagram and the BAO distances, respectively.
\begin{table}[htbp]
\begin{center}
\scalebox{1.2}{
\begin{tabular}{|c|ccc|}
\hline
& $\Lambda$CDM & Full galileon & Cubic galileon \\
\hline
\rule{0pt}{2.6ex} $\Omega_b h^{2}$ & 0.02228 $\pm$ 0.00014 & 0.02197 $\pm$ 0.00013 & 0.02189 $\pm$ 0.00013 \\
$\Omega_c h^{2}$ & 0.1187 $\pm$ 0.0010 & 0.1231 $\pm$ 0.0010 & 0.12429 $\pm$ 0.00097 \\
$100\theta_{MC}$ & 1.04091 $\pm$ 0.00029 & 1.04051 $\pm$ 0.00029 & 1.04075 $\pm$ 0.00037 \\
$\tau$ & 0.069 $\pm$ 0.012 & 0.0054 $\pm$ 0.0041 & 0.0051 $\pm$ 0.0036 \\
$\text{ln}(10^{10}A_{s})$ & 3.070 $\pm$ 0.023 & 2.9478 $\pm$ 0.0094 & 2.9481 $\pm$ 0.0089 \\
$n_{s}$ & 0.9662 $\pm$ 0.0040 & 0.9564 $\pm$ 0.0037 & 0.9575 $\pm$ 0.0035 \\
$c_{2}$ & - & -7.30 $\pm$ 0.88 & -4.454 $\pm$ 0.029 \\
$c_{3}$ & - & -2.54 $\pm$ 0.60 & -0.7420 $\pm$ 0.0047 \\
$c_{4}$ & - & -0.65 $\pm$ 0.25 & - \\
$c_{5}$ & - & -0.29 $\pm$ 0.13 & - \\
$c_{G}$ & - & 0.0100 $\pm$ 0.0071 & - \\
$H_{0}$ & 67.73 $\pm$ 0.63 & 75.24 $\pm$ 0.49 & 75.53 $\pm$ 0.88 \\
$\sigma_{8}$ & 0.8180 $\pm$ 0.0086 & 0.904 $\pm$ 0.011 & 0.8988 $\pm$ 0.0046 \\
$\Omega_{\varphi}^{*}/\Omega_{m}^{*}$ & - & 0.000084 $\pm$ 0.000070 & 0.0024 $\pm$ 0.0015 \\
$r_{d}$ & 147.54 $\pm$ 0.24 & 146.73 $\pm$ 0.24 & 146.34 $\pm$ 0.29 \\
$z_{rei}$ & 9.1 $\pm$ 1.1 & 1.07 $\pm$ 0.88 & 1.01 $\pm$ 0.80 \\
$10^{9}A_{s}e^{-2\tau}$ & 1.875 $\pm$ 0.011 & 1.886 $\pm$ 0.011 & 1.888 $\pm$ 0.011 \\
\hline
\rule{0pt}{2.6ex} $\chi^{2} (\text{CMB})$ & 12946 & 12966 & 12993 \\
$\chi^{2} (\text{BAO})$ & 5.6 & 30.4 & 29.9 \\
$\chi^{2} (\text{JLA})$ & 706.7 & 723.3 & 723.6 \\
\hline
\end{tabular}
}
\caption{Constraints on the free parameters and a few derived parameters for the three base models from CMB, BAO and JLA combined data. $\chi^{2}$ values given in the table are contributions from each probe to the $\chi^{2}$ of the global best-fit.}
\label{table:Contraintes CMB+BAO+JLA}
\end{center}
\end{table}
\newline
\newline
As can be seen in Table \ref{table:separate Chi2} that shows $\chi^{2}$ values of the best-fits to each dataset separately, the $\Lambda$CDM and Full galileon models are able to fit the three datasets correctly, while the Cubic galileon has difficulties to fit CMB data properly, especially the ISW part (low-$\ell$) of the temperature power spectrum, and, to a lesser extent, the lensing power spectrum (see Figure \ref{fig:CMB+BAO+JLA best-fit CMB}).
\begin{table}[htbp]
\begin{center}
\scalebox{1.2}{
\begin{tabular}{|c|ccc|}
\hline
& $\Lambda$CDM & Full galileon & Cubic galileon \\
\hline
\rule{0pt}{2.6ex} $\chi^{2}_{\text{CMB}}$ & 12943 & 12949 & 12976 \\
$\chi^{2}_{\text{BAO}}$ & 3.0 & 2.5 & 2.5 \\
$\chi^{2}_{\text{JLA}}$ & 695.1 & 695.6 & 695.7 \\
\hline
\end{tabular}
}
\caption{$\chi^{2}$ values of the best-fit to each dataset separately for the three base models.}
\label{table:separate Chi2}
\end{center}
\end{table}
However, only the $\Lambda$CDM model is able to fit correctly the combination of CMB, BAO and JLA data. Indeed, the contributions from each probe to the total $\chi^{2}$ are higher for the Galileon models than for the $\Lambda$CDM model (see Table \ref{table:Contraintes CMB+BAO+JLA}). This is not so significant in the case of JLA and CMB data given the high number of measurements. On the other hand, the Galileon models have $\chi^{2} \left( \text{BAO} \right) \sim 30$ for only 8 data points, significantly larger than what is seen in the $\Lambda$CDM model. Thus, the Galileon base models appear to be in tension with cosmological observations, especially with BAO measurements (see Figure \ref{fig:CMB+BAO+JLA best-fit BAO}). Note also that the Cubic galileon model provides a worse fit to data than the Full galileon model, as indicated by higher CMB $\chi^{2}$ values in the Cubic model but also by non-zero Full galileon model best-fit values for $c_4, c_5$ and (to a lesser extent) $c_G$ (see Table \ref{table:Contraintes CMB+BAO+JLA}). \\
\newline
Another interesting feature of the Galileon models compared to $\Lambda$CDM is that CMB data tends to favour very low values of the optical depth at reionization, $\tau$, and thus low values of the normalization factor of primordial perturbations, $A_{s}$, since most of CMB observables 
depend on the combination $A_{s}e^{-2\tau}$ (see Table \ref{table:Contraintes CMB+BAO+JLA} and Figure \ref{fig:CMB+BAO+JLA usual}). In fact, only the lensing spectrum and the low multipoles of the TE spectrum slightly break the degeneracy. In the case of the Galileon models, the two effects 
pull the value of $\tau$ (or equivalently of $A_{s}$) in opposite directions. Polarization favours reasonably high values of $\tau \sim 0.06$ while the normalization of the lensing power spectrum favours very low values of $\tau \sim 0.005$. This comes from the fact that the gravitational potential is deeper in the selected Galileon scenarios compared to $\Lambda$CDM, which leads to more lensing. The very low value predicted for $\tau$ in the Galileon models has important consequences since it sets the value of the reionization redshift, at which most of the Universe is reionized, to $z_{rei} \sim 1$. On the other hand, this redshift can be constrained by direct astrophysical observations \cite{Bouwens-2015}, and is found to be $z_{rei} \gtrsim 6$, leading to another incompatibility. In fact, among all Galileon scenarios saved in the chains, not a single one predicts a $z_{rei}$ value above 6. This is the final nail in the coffin of Galileon base models. \\
\newline
It may, however, be possible to alleviate the two incompatibilities previously encountered by adding one of the available additional parameters, 
as will be presented in the following section.

\subsection{Extended models}
\label{subsection:subsection5.2}

The goal of this section is to check whether adding one more parameter to the base models, with proper justification, could solve the incompatibility on the reionization redshift and the tension with BAO measurements. While it is hard to explain why Galileon models fail to reproduce CMB, BAO and JLA measurements simultaneously, we have seen that the reason why Galileon models predict a very low value for $\tau$, and thus $z_{rei}$, is related to the higher lensing effect in the selected Galileon galileon scenarios. Therefore, we will consider in the following additional parameters that are likely to impact the lensing power spectrum, and will check whether it also solves the BAO tension.

\subsubsection{Lensing normalization}
\label{subsubsection:subsubsection5.2.1}

Gravitational lensing by large-scale structures impacts the CMB temperature and polarization. Acoustic peaks and troughs in the TT, TE and EE spectra are smoothed, some polarization E-modes are converted to B-modes and significant non-Gaussianity is generated in the form of a non-zero connected 4-point correlation function. The lensing power spectrum $C_{\ell}^{\phi\phi}$ measured by the Planck collaboration corresponds to the latter effect and is determined from lens reconstructions using 4-point correlation functions of both temperature and polarization data \cite{Ade-2013, Ade-2015}. 
That direct measurement and the imprint of lensing embedded in the CMB anisotropy power spectra (in temperature and polarization) are expected to be consistent -- within the considered theoretical model.
To assess this, the Planck collaboration introduced the $A_L$ parameter in their cosmological fits that scales the theoretical prediction for the lensing power spectrum in each point of the parameter space and is also used to predict the lensed CMB spectra. The theoretical expectation is $A_L =1$. However, Planck data indicate a preference for $A_L > 1$ in the $\Lambda$CDM model, more significant when only CMB spectra are used, e.g. $A_L=1.22 \pm 0.11$ from fits using TT and low multipole polarization 2015 data \cite{Planck-2015}. Planck direct lensing reconstruction pulls the constraint towards $A_L=1$ without restoring full agreement  (for more complete and quantitative results, see the final publication \cite{Planck-2018}). If this tension is not purely of statistical origin, it may be the hint of residual systematics or indicate a breakdown of the theoretical model. It is therefore relevant to include $A_L$ in our study.
\begin{figure}[htbp]
\begin{center}
\includegraphics[width=\textwidth]{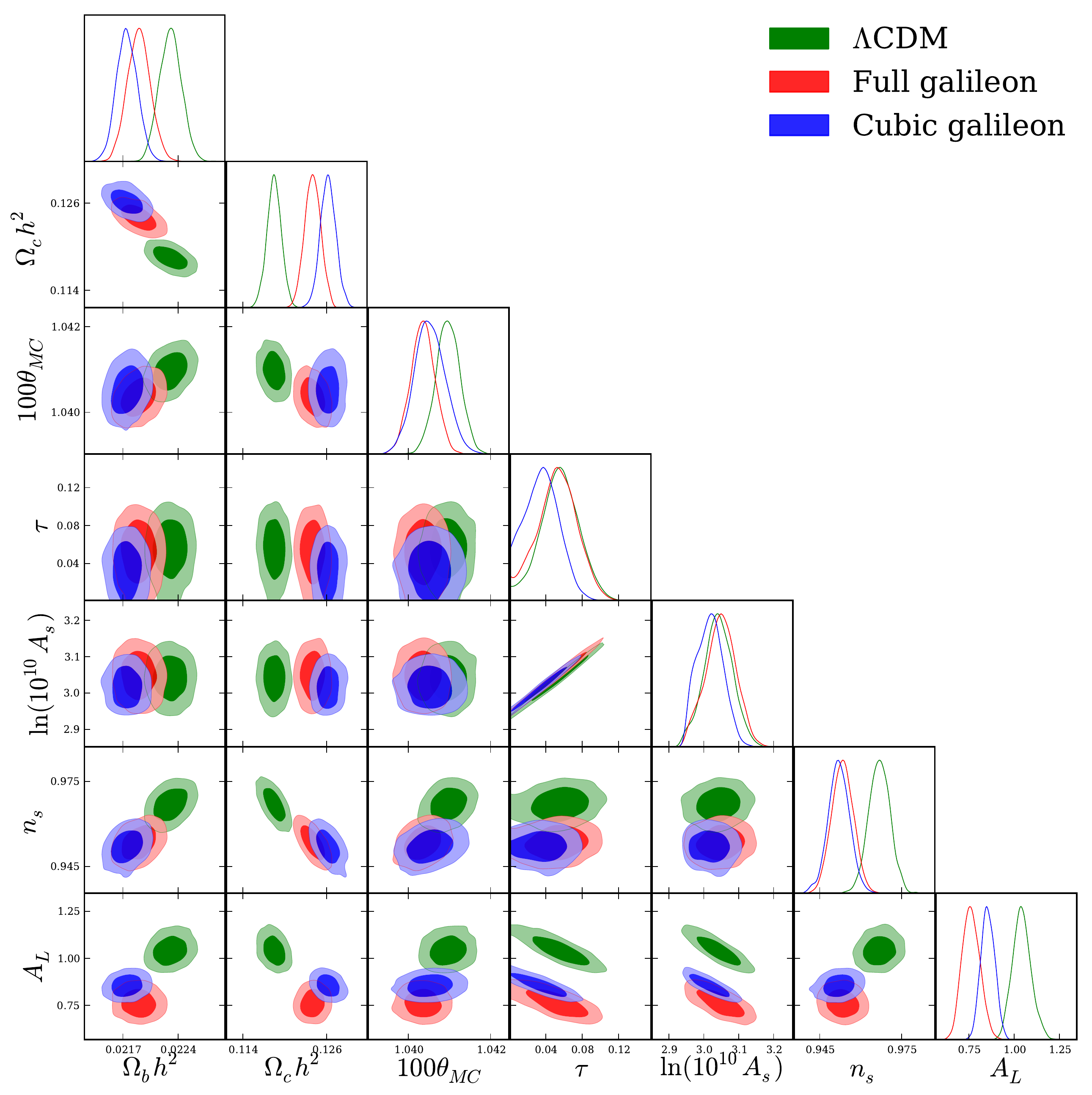}
\caption{Marginalized 1D and 2D constraints for all parameters common to the three models extended to $A_{L}$, from fits to CMB, BAO and JLA combined data.}
\label{fig:CMB+BAO+JLA_Al usual}
\end{center}
\end{figure}
\begin{table}[htbp]
\begin{center}
\scalebox{1.2}{
\begin{tabular}{|c|ccc|}
\hline
& $\Lambda$CDM & Full galileon & Cubic galileon \\
\hline
\rule{0pt}{2.6ex} $\Omega_b h^{2}$ & 0.02230 $\pm$ 0.00014 & 0.02190 $\pm$ 0.00015 & 0.02175 $\pm$ 0.00013 \\
$\Omega_c h^{2}$ & 0.1185 $\pm$ 0.0011 & 0.1240 $\pm$ 0.0012 & 0.1262 $\pm$ 0.0011 \\
$100\theta_{MC}$ & 1.04096 $\pm$ 0.00030 & 1.04035 $\pm$ 0.00030 & 1.04052 $\pm$ 0.00038 \\
$\tau$ & 0.055 $\pm$ 0.021 & 0.052 $\pm$ 0.027 & 0.036 $\pm$ 0.018 \\
$\text{ln}(10^{10}A_{s})$ & 3.041 $\pm$ 0.042 & 3.048 $\pm$ 0.044 & 3.018 $\pm$ 0.037 \\
$n_{s}$ & 0.9668 $\pm$ 0.0040 & 0.9534 $\pm$ 0.0040 & 0.9520 $\pm$ 0.0040 \\
$A_{L}$ & 1.043 $\pm$ 0.052 & 0.763 $\pm$ 0.049 & 0.855 $\pm$ 0.039 \\
$c_{2}$ & - & -7.74 $\pm$ 0.69 & -4.391 $\pm$ 0.035 \\
$c_{3}$ & - & -2.87 $\pm$ 0.47 & -0.7314 $\pm$ 0.0059 \\
$c_{4}$ & - & -0.79 $\pm$ 0.19 & - \\
$c_{5}$ & - & -0.320 $\pm$ 0.090 & - \\
$c_{G}$ & - & 0.022 $\pm$ 0.010 & - \\
$H_{0}$ & 67.86 $\pm$ 0.48 & 73.68 $\pm$ 0.59 & 74.50 $\pm$ 0.60 \\
$\sigma_{8}$ & 0.806 $\pm$ 0.017 & 0.987 $\pm$ 0.027 & 0.935 $\pm$ 0.018 \\
$\Omega_{\varphi}^{*}/\Omega_{m}^{*}$ & - & 0.00047 $\pm$ 0.00022 & 0.0024 $\pm$ 0.0015 \\
$r_{d}$ & 147.57 $\pm$ 0.24 & 146.56 $\pm$ 0.25 & 146.00 $\pm$ 0.31 \\
$z_{rei}$ & 7.6 $\pm$ 2.2 & 7.4 $\pm$ 2.4 & 5.62 $\pm$ 2.32 \\
$10^{9}A_{s}e^{-2\tau}$ & 1.873 $\pm$ 0.011 & 1.898 $\pm$ 0.011 & 1.903 $\pm$ 0.011 \\
\hline
\rule{0pt}{2.6ex} $\chi^{2} (\text{CMB})$ & 12945 & 12960 & 12984 \\
$\chi^{2} (\text{BAO})$ & 5.2 & 18.4 & 22.5 \\
$\chi^{2} (\text{JLA})$ & 706.6 & 718.9 & 721.6 \\
\hline
\end{tabular}
}
\caption{Constraints on the free parameters and a few derived parameters for the three models extended to $A_{L}$ from CMB, BAO and JLA combined data. $\chi^{2}$ values given in the table are contributions from each probe to the $\chi^{2}$ of the global best-fit.}
\label{table:Contraintes CMB+BAO+JLA_Al}
\end{center}
\end{table}
\newline
\newline
Figure \ref{fig:CMB+BAO+JLA_Al usual} shows the contours and marginalized likelihoods obtained using the combination of CMB, BAO and JLA data for the parameters common to the three models extended to $A_{L}$. The confidence intervals for each parameters are summarized in Table \ref{table:Contraintes CMB+BAO+JLA_Al}. As expected since $A_{L}$ plays directly on the normalization of the lensing power spectrum, it is highly correlated with $A_{s}$ and therefore with $\tau$. It is possible to keep reasonable values of $\tau$ and still fit the lensing power spectrum correctly in the Galileon models by lowering its normalization, i.e. by having lower values for $A_{L}$. This way, the predicted values for $\tau$ and $z_{rei}$ become similar in Galileon models and $\Lambda$CDM model and are compatible with direct observations. \\
\newline
However, though adding this new parameters solves the issue with reionization, it does not solve the tension with BAO measurements. Despite the lower $\chi^{2} \left( BAO \right)$ for the best-fits of the two Galileon models, which is to be expected when adding a new parameter, they are still at some large and unlikely value. So, Galileon models remain in tension with BAO measurements in this first extension.

\subsubsection{Active neutrinos masses}
\label{subsubsection:subsubsection5.2.2}

\begin{figure}[htbp]
\begin{center}
\includegraphics[width=\textwidth]{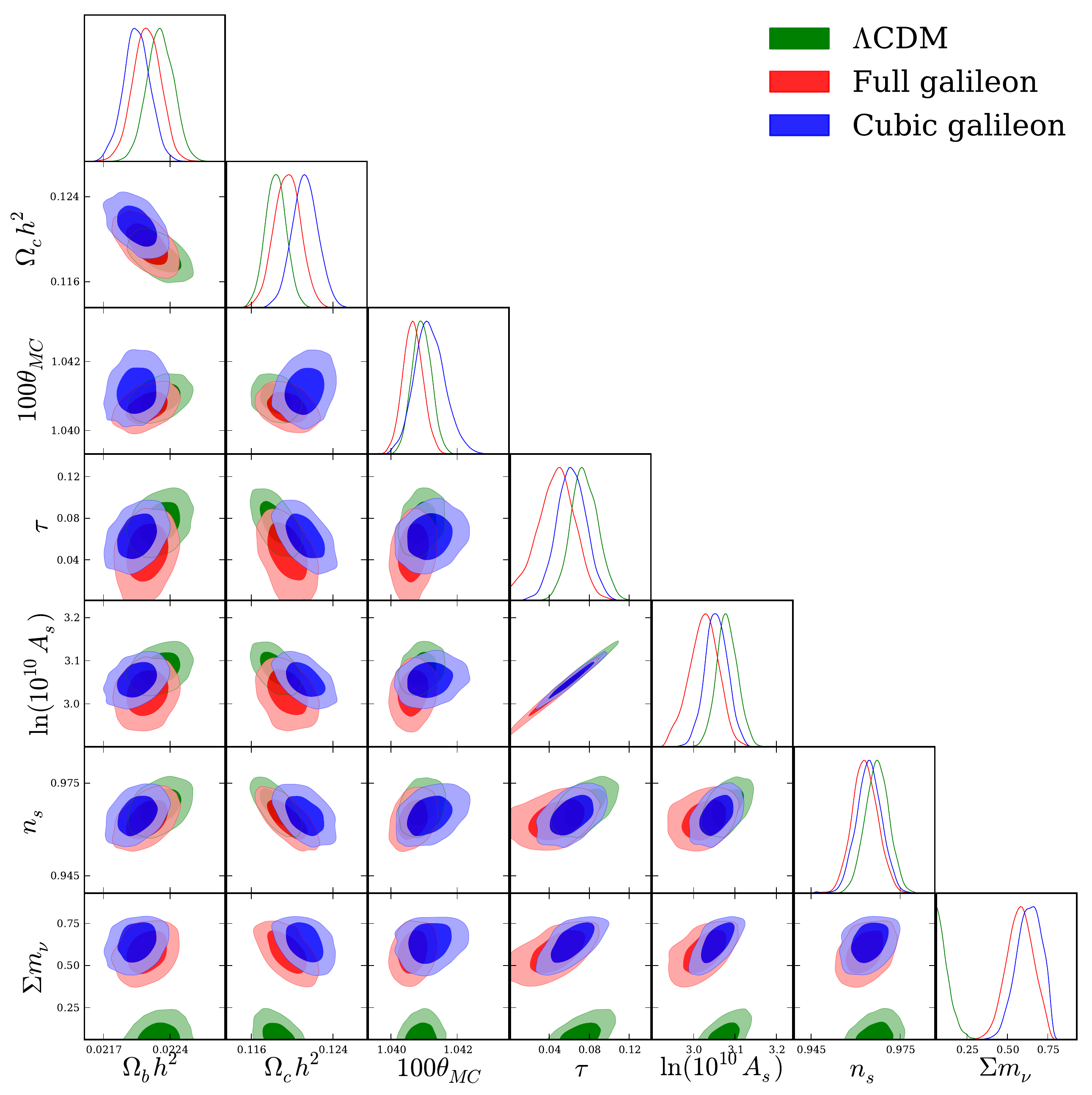}
\caption{Marginalized 1D and 2D constraints for all parameters common to the three models extended to $\sum m_{\nu}$, from fits to CMB, BAO and JLA combined data.}
\label{fig:CMB+BAO+JLA_mnu usual}
\end{center}
\end{figure}

\begin{figure}[htbp]
\begin{center}
\includegraphics[width=0.7\textwidth]{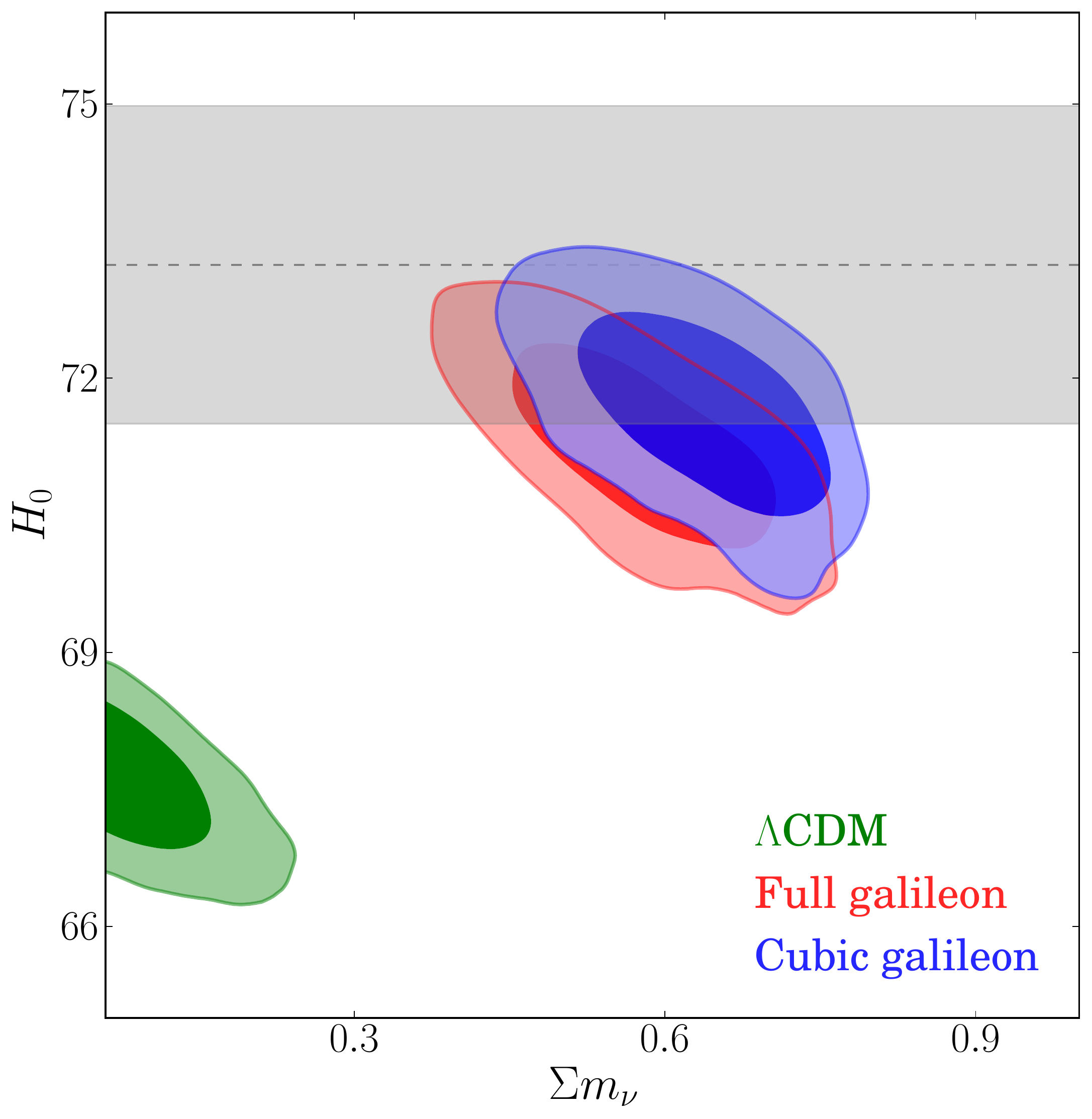}
\caption{2D contours in the $\left( \sum m_{\nu}, H_{0} \right)$ plane for the three considered models from the fit to CMB, JLA and BAO combined data. The grey band corresponds to the 68\% confidence limit of the direct measurement of $H_{0}$ from \cite{Riess-2016}.}
\label{fig:H0 mnu}
\end{center}
\end{figure}

The second relevant free parameter that has a significant effect on CMB lensing is the sum of active neutrinos masses, $\sum m_{\nu}$, as was already explored in \cite{Barreira-2014}. 
The three active neutrinos of the standard model of particle physics carry a mass, as proved by the observation of neutrino oscillations \cite{Kajita-1998}, with the present limits:
\begin{equation}
    0.06 \ \text{eV} < \sum m_{\nu} < 6.6 \ \text{eV}
\end{equation}
The lower bound comes from neutrino oscillations \cite{Kraus-2004} and the upper bound from direct experiments \cite{Aseev-2011}. There is currently no evidence that $\sum m_{\nu} = 0.06 \ \text{eV}$ as was assumed in the base models, so it is perfectly relevant to let this parameter free. Moreover, in the above range of masses, neutrinos do not modify significantly the shape of the CMB power spectra except for the lensing power spectrum and large scales of the temperature power spectrum. More massive neutrinos interact more via gravitation with CDM and baryons, but since they do not enter structure formation because of free streaming, this interaction slows down the infall of matter inside gravitational wells. So, larger neutrino masses mean lower density contrast which lowers the gravitational potential and the lensing effect. \\
\newline
Figure \ref{fig:CMB+BAO+JLA_mnu usual} shows the contours and marginalized likelihoods obtained using the combination of CMB, BAO and JLA data for the parameters common to the three models extended to $\sum m_{\nu}$. The confidence intervals for each parameters are summarized in Table \ref{table:Contraintes CMB+BAO+JLA_mnu}.
\begin{table}[htbp]
\begin{center}
\scalebox{1.2}{
\begin{tabular}{|c|ccc|}
\hline
& $\Lambda$CDM & Full galileon & Cubic galileon \\
\hline
\rule{0pt}{2.6ex} $\Omega_b h^{2}$ & 0.02230 $\pm$ 0.00014 & 0.02216 $\pm$ 0.00015 & 0.02205 $\pm$ 0.00014 \\
$\Omega_c h^{2}$ & 0.1184 $\pm$ 0.0010 & 0.1195 $\pm$ 0.0013 & 0.1213 $\pm$ 0.0013 \\
$100\theta_{MC}$ & 1.04093 $\pm$ 0.00029 & 1.04066 $\pm$ 0.00030 & 1.04118 $\pm$ 0.00046 \\
$\tau$ & 0.075 $\pm$ 0.014 & 0.0047 $\pm$ 0.018 & 0.062 $\pm$ 0.015 \\
$\text{ln}(10^{10}A_{s})$ & 3.081 $\pm$ 0.026 & 3.024 $\pm$ 0.035 & 3.055 $\pm$ 0.027 \\
$n_{s}$ & 0.9673 $\pm$ 0.0040 & 0.9632 $\pm$ 0.0043 & 0.9645 $\pm$ 0.0044 \\
$\sum m_{\nu}$ & 0.115 $\pm$ 0.045 & 0.578 $\pm$ 0.083 & 0.634 $\pm$ 0.077 \\
$c_{2}$ & - & -7.30 $\pm$ 0.96 & -4.243 $\pm$ 0.045 \\
$c_{3}$ & - & -2.64 $\pm$ 0.63 & -0.7066 $\pm$ 0.0076 \\
$c_{4}$ & - & -0.70 $\pm$ 0.25 & - \\
$c_{5}$ & - & -0.28 $\pm$ 0.12 & - \\
$c_{G}$ & - & 0.017 $\pm$ 0.013 & - \\
$H_{0}$ & 67.43 $\pm$ 0.52 & 71.25 $\pm$ 0.75 & 71.66 $\pm$ 0.76 \\
$\sigma_{8}$ & 0.811 $\pm$ 0.011 & 0.806 $\pm$ 0.018 & 0.793 $\pm$ 0.015 \\
$\Omega_{\varphi}^{*}/\Omega_{m}^{*}$ & - & 0.00020 $\pm$ 0.00016 & 0.0047 $\pm$ 0.0026 \\
$r_{d}$ & 147.59 $\pm$ 0.24 & 147.25 $\pm$ 0.27 & 146.52 $\pm$ 0.41 \\
$z_{rei}$ & 9.7 $\pm$ 1.2 & 6.8 $\pm$ 2.1 & 8.6 $\pm$ 1.5 \\
$10^{9}A_{s}e^{-2\tau}$ & 1.874 $\pm$ 0.011 & 1.874 $\pm$ 0.012 & 1.874 $\pm$ 0.011 \\
\hline
\rule{0pt}{2.6ex} $\chi^{2} (\text{CMB})$ & 12946 & 12947 & 12961 \\
$\chi^{2} (\text{BAO})$ & 5.5 & 16.9 & 18.5 \\
$\chi^{2} (\text{JLA})$ & 706.7 & 718.3 & 716.2 \\
\hline
\end{tabular}
}
\caption{Constraints on the free parameters and a few derived parameters for the three models extended to $\sum m_{\nu}$ from CMB, BAO and JLA combined data. $\chi^{2}$ values given in the table are contributions from each probe to the $\chi^{2}$ of the global best-fit.}
\label{table:Contraintes CMB+BAO+JLA_mnu}
\end{center}
\end{table}
The conclusions are basically the same as for the previous extension, as expected since they have a similar effect on the CMB power spectra and do not change much the predictions of BAO scales. The incompatibility with reionization is solved by increasing $\sum m_{\nu}$, but the BAO tension remains. \\
\newline
It is interesting to note that, contrary to $\Lambda$CDM, the best fitting Galileon scenarios predict a non-zero sum of neutrino masses, as was noticed also in other modified gravity models (see e.g. \cite{Barreira-2015,Bellomo-2017,Baldi-2013,Dirian-2017}). The sum of neutrino masses is strongly correlated with $\Omega_{m}$ and the Hubble constant, whose constraint from Planck \cite{Planck-2015,Planck-2018} in $\Lambda$CDM is in tension with the direct measurement from \cite{Riess-2016}. 
In $\Lambda$CDM, the higher $\sum m_{\nu}$, the lower $H_{0}$. This is also the case for the Galileon models, as can be seen in Figure \ref{fig:H0 mnu} (see also Tables \ref{table:Contraintes CMB+BAO+JLA},\ref{table:Contraintes CMB+BAO+JLA_mnu}). Altogether the best fitting Galileon scenarios predict a non-zero $\sum m_{\nu}$ compatible with particle physics bounds and a high value of $H_{0}$ in agreement with the direct measurement of \cite{Riess-2016}. 
To be fully satisfactory, modified gravity models should also relieve the tension between CMB and clustering data on $\sigma_8$ (late-time clustering amplitude) present in $\Lambda$CDM \cite{Planck-2015,Planck-2018} and Galileon models (see Table \ref{table:Contraintes CMB+BAO+JLA_mnu}). Note, however, that $\sigma_{8}$, $H_{0}$ and $\sum m_{\nu}$ are correlated and degenerate with modified gravity effects, as discussed in \cite{Planck-2016,Planck-2018}. At present, tensions on these parameters are not strong enough to indicate a breakdown of $\Lambda$CDM. Until a standard explanation for these tensions is found, they remain a good reason to keep on looking at gravity models.

\subsection{Constraints from Gravitational Waves}
\label{subsection:subsection5.3}

\begin{figure}[htbp]
\begin{center}
\includegraphics[width=0.7\textwidth]{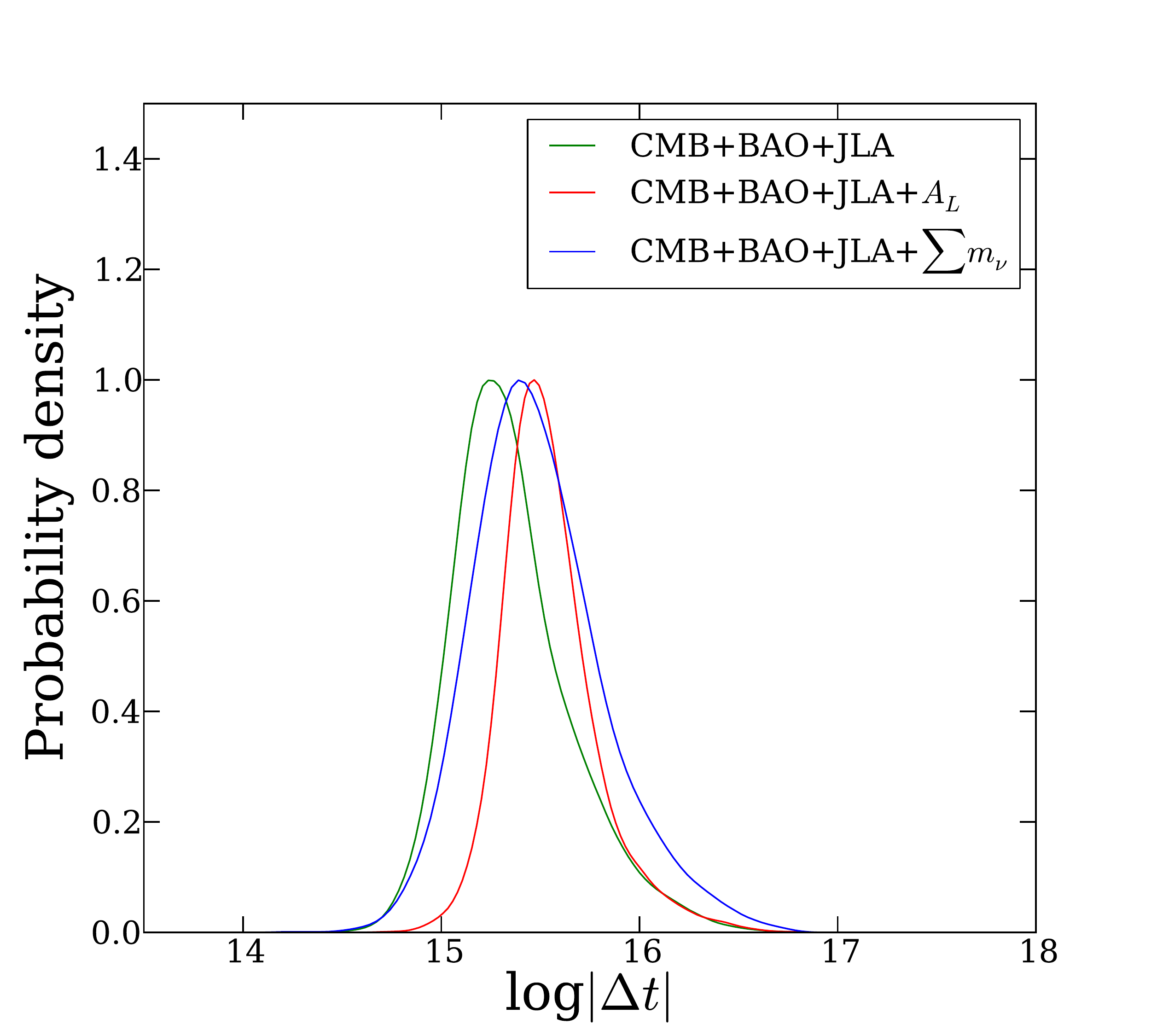}
\caption{Predictions for the time delay, log$|\Delta t|$, between the arrival of the GW signal and light emitted by the neutron star binary merger described in section \ref{subsection:subsection3.4}, for all scenarios selected by the Markov chains of the three Full galileon cosmological fits.}
\label{fig:logdt}
\end{center}
\end{figure}

One originality of this work is to treat the time delay $\Delta t$ between gravitational waves and the electromagnetic counterpart of GW170817 as a derived parameter that can be computed \textit{a posteriori} for every scenario selected by the Markov chains. The constraints on $\Delta t$ derived from CMB, BAO and JLA data can thus be confronted with the measurement $\Delta t = 1.74 \pm 0.05 \ \text{s}$. \\
\newline
As explained previously, the $\Lambda$CDM and Cubic galileon models both predict that gravitational waves speed is equal to the speed of light. It implies that $|\Delta t| = |\delta t| < 10 \ \text{s}$ which is, by construction, compatible with GW170817. So, the constraints from CMB, BAO and JLA on $\Delta t$ are only relevant for the Full galileon model. The likelihood distributions of log$|\Delta t|$ for the Full galileon base model and its two extensions are presented in Figure \ref{fig:logdt}.
Full galileon scenarios selected by the Markov chains of the fit to all cosmological datasets combined predict a huge delay between gravitational waves and light, of the order of:
\begin{equation}
    |\Delta t| > 10^{14} \ \text{sec} \sim \ \text{a few million years}
\end{equation}
This prediction, consistent with the results presented in \cite{Ezquiaga-2017,Wang-2017,Sakstein-2017}, is totally incompatible with the observation of GW170817, which definitely rules out the Full galileon as a viable cosmological model.

\section{Discussion}
\label{section:section6}

\subsection{Tensions with cosmological observations}
\label{subsection:subsection6.1}

\begin{figure}[htbp]
\begin{center}
\includegraphics[width=\textwidth]{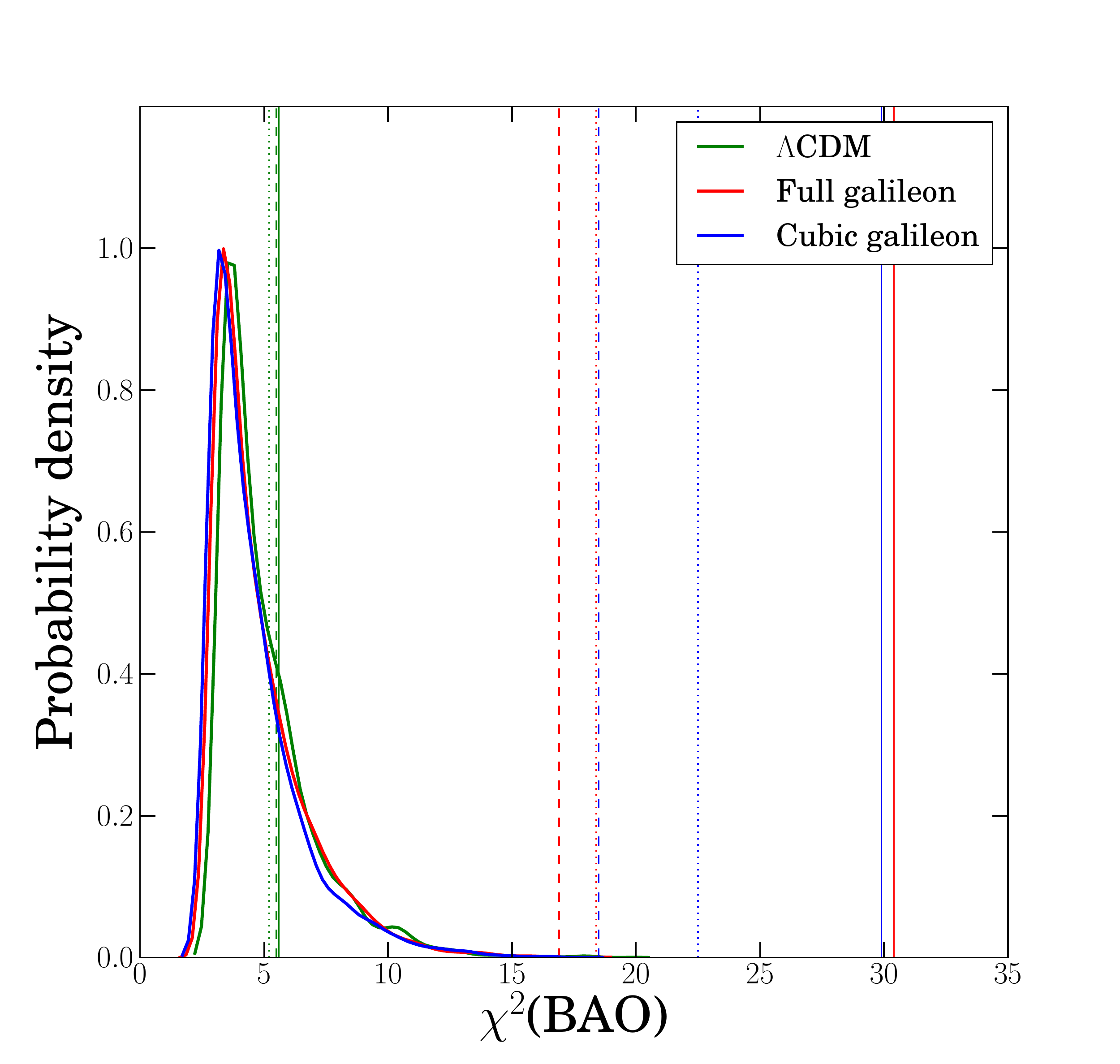}
\caption{$\chi^{2}$ distributions of points selected in the Markov Chains of the base model fits to BAO data only. Vertical lines indicate the best-fit $\chi^{2}$ values of fits to all data combined for the base models (full lines), and their extensions to $A_{L}$ (dotted lines) and $\sum m_{\nu}$ (dashed lines).}
\label{fig:chisq_bao}
\end{center}
\end{figure}

The previous section shows that the Galileon base and extended models have serious difficulties to reproduce BAO measurements when fitting to CMB, BAO and JLA data simultaneously. To understand why this is so, we remind that BAO measurements are sensitive to the combination of parameters $P = c/ \left( H_{0}r_{d} \right)$, where $r_{d}$ is the sound horizon at the drag epoch, given by :
\begin{equation}
    r_{d} = \int_{z_{d}}^{+\infty} \frac{c_{s} \left( z \right)}{H \left( z \right)} dz
\end{equation}
with $c_{s}$ the sound speed in the baryon-photon fluid. The sound horizon depends only on the physics of the baryon-photon plasma until the end of recombination. So, $r_{d}$ is expected to be similar in models that have similar early universe behaviour, in our case if dark energy is still negligible until the drag epoch. The values of $\Omega_{\varphi}^{*}/\Omega_{m}^{*}$ (ratio of the Galileon and matter energy densities at recombination) in Tables \ref{table:Contraintes CMB+BAO+JLA}, \ref{table:Contraintes CMB+BAO+JLA_Al} and \ref{table:Contraintes CMB+BAO+JLA_mnu} show that this is indeed the case for the selected Galileon scenarios. \\
\newline
In the absence of dark energy, the value of $r_{d}$ depends non trivially on $\lbrace \Omega_{b}, \Omega_{c}, H_{0}\rbrace$. Thus, in our parameterization where $H_{0}$ is computed from the value of $\theta_{MC}$, BAO predictions depend on $\lbrace \Omega_{b}h^{2}, \Omega_{c}h^{2}, 100 \theta_{\text{MC}} \rbrace$. However, these parameters are very well constrained by CMB data, through the positions and amplitudes of the temperature power spectrum peaks. These features depend mostly on the dynamics of the baryon-photon plasma at recombination and are fairly model independent if dark energy is negligible at that epoch. In that case, when CMB data is used, there is not much freedom left to fit BAO measurements. \\
\newline
Indeed, at given values of $\Omega_{b}h^{2}$ and $\Omega_{c}h^{2}$, fixing the Hubble constant is equivalent to fixing $\theta_{\text{MC}}$ whose prediction depends strongly on the model through the comoving angular distance $D_{M}$ which is determined by the late time history of the Universe:
\begin{equation}
    \theta_{MC} = \frac{r_{s}}{D_{M} \left( z_{*} \right)} = \frac{\int_{z_{*}}^{+\infty} \frac{c_{s} \left( z \right)}{H \left( z \right)}dz}{\int_{0}^{z_{*}} \frac{1}{H \left( z \right)}dz}
\end{equation}
Here $z_{*}$ is the redshift at recombination and $r_{s}$ is the sound horizon at recombination, which is very similar to $r_{d}$. Tight constraints on $\Omega_{b}h^{2}$, $\Omega_{c}h^{2}$ and $100 \theta_{\text{MC}}$ from CMB data will thus favour model parameters that give approximately the same $D_{M} \left( z_{*} \right)$ as in  $\Lambda$CDM. In Galileon fits, the parameters that enter the  prediction of $D_{M} \left( z_{*} \right)$ are the Galileon $c_i$ parameters, $\Omega_{m}$ and $H_{0}$, with the constraint that $\Omega_{m}h^{2}$ be set by CMB data. 
So, the disagreement between Galilon models and data observed in combined fits means that the region favoured by CMB data does not overlap the region favoured by BAO data. 
In other words, there is a tension between the early universe CMB data and the late time BAO data, concerning in particular the value of $\Omega_{m}$. Indeed the expansion rate is higher in the Galileon model than in the $\Lambda$CDM model (see Tables \ref{table:Contraintes CMB+BAO+JLA}, \ref{table:Contraintes CMB+BAO+JLA_Al} and \ref{table:Contraintes CMB+BAO+JLA_mnu}). As CMB data constrains $\Omega_{m}h^{2}$ to remain the same, a higher $H_0$ implies a lower $\Omega_m$. This creates an impossibility for the Galileon models to satisfy both the constraint from CMB data on $D_{M} \left( z_{*} \right)$ to be close to the $\Lambda$CDM value when dark energy is subdominant, and the constraints on $H(z)$ and $D_A(z)$ from BAO, where the Galileon field dominates. \\ 
\newline
To quantify rigorously the disagreement between data and predictions from the Galileon model would require an extensive statistical analysis, which is not reasonably conceivable for such a model. It is possible, though, to have an idea of this disagreement by comparing $\chi^{2} \left( \text{BAO} \right)$, the BAO contribution to the global $\chi^2$ from the best combined fits, with the $\chi^2$ distributions from the fits to BAO data only (see Figure \ref{fig:chisq_bao}).
A model in agreement with BAO data would have its $\chi^{2} \left( \text{BAO} \right)$ from the best combined fit near the peak of the $\chi^{2}$ distribution, as is the case for the $\Lambda$CDM model. The fact that the $\chi^{2} \left( \text{BAO} \right)$ values from the best fits of the Galileon models are so far in the tail of the $\chi^{2}$ distributions illustrates the disagreement between BAO data and the other datasets. It is possible to quantify the disagreement, in order of magnitude, based on these $\chi^{2}$ distributions. This is done by computing the p-value of the best combined fit. When no point with such high $\chi^{2}$ where selected in the fits to BAO data only, we estimate the p-value to be $< 3/N_{\text{chains}}$, where $N_{\text{chains}}$ is the number of points selected in the Markov chains. The Full galileon models are found to be in tension with BAO data at the level of $\sim 3.5 \sigma$ (base), $\sim 3.8 \sigma$ (extension to $\sum m_{\nu}$), $> 3.9 \sigma$ (extension to $A_{L}$). Cubic galileon models are found to be in $> 3.8 \sigma$ (all) tension with BAO data. This result completes and extends the conclusion, from previous works \cite{Renk-2017,Peirone-2017}, that the tracker solutions of the Galileon models were ruled out by cosmological data.

\subsection{Tracker solutions}
\label{subsection:subsection6.2}

\begin{figure}[htbp]
\begin{center}
\begin{subfigure}{0.5\textwidth}
  \centering
  \includegraphics[width=\linewidth]{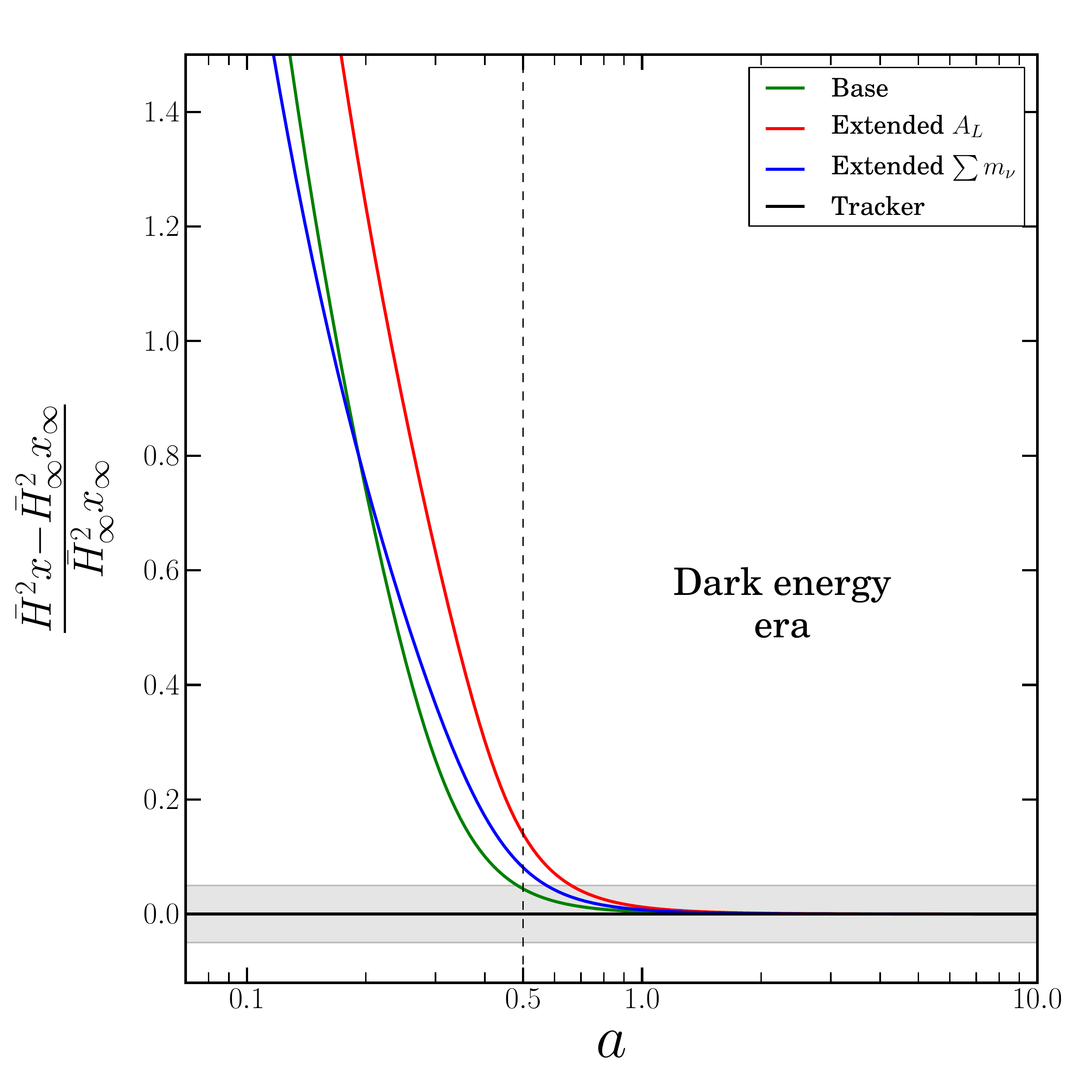}
\end{subfigure}%
\begin{subfigure}{0.5\textwidth}
  \centering
  \includegraphics[width=\linewidth]{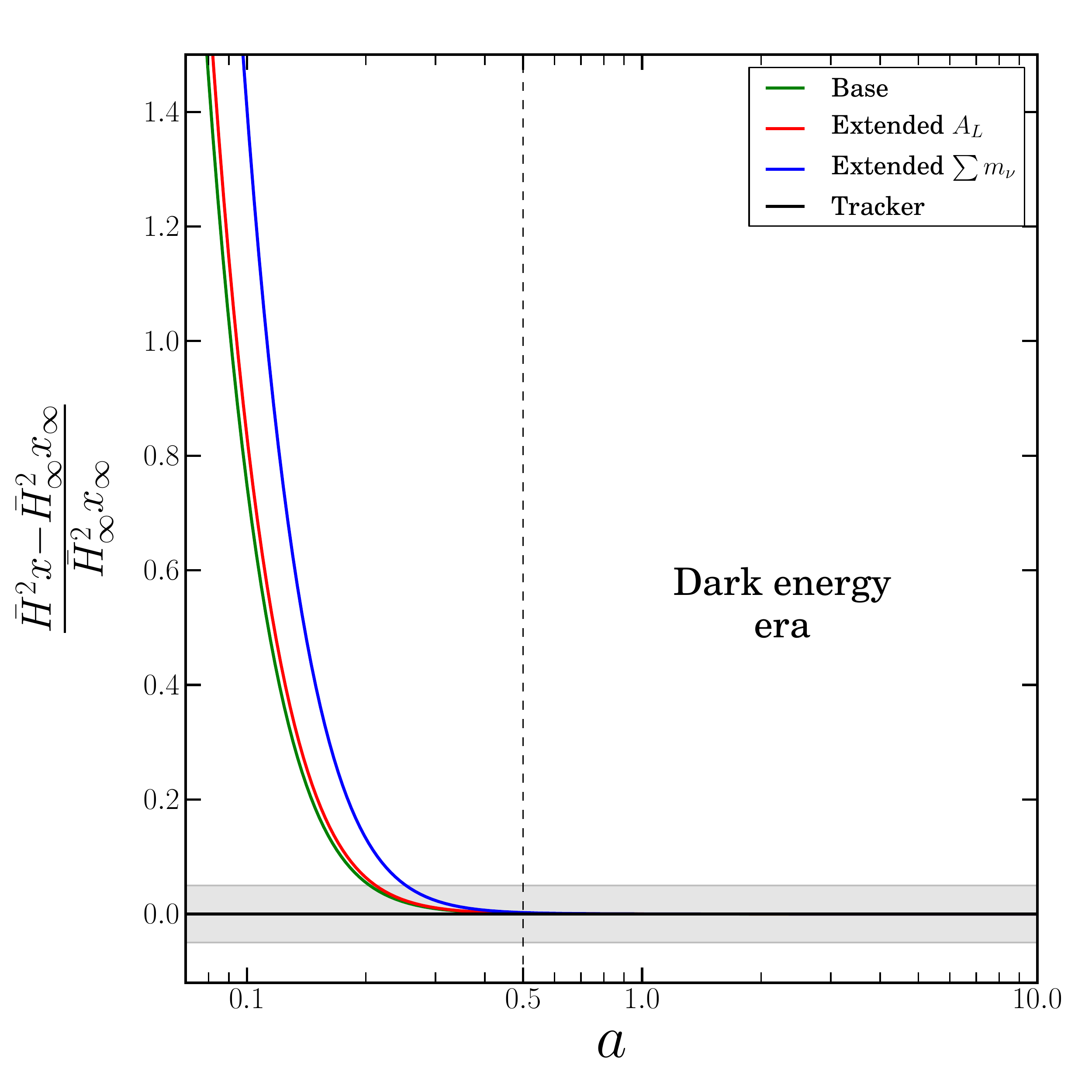}
\end{subfigure}
\end{center}
\caption{Evolution of $\Hbar^{2} x$ as a function of the scale factor for the best fits of the Full galileon (left) and Cubic galileon (right) base and extended models. The tracker solutions are located on the horizontal axis. The grey bands correspond to a gap of 5\% with respect to the associated tracker.}
\label{fig:trackers}
\end{figure}

One of the main goals of this work was to confirm or invalidate results from previous studies \cite{Barreira-2014,Barreira-2014b,Renk-2017,Peirone-2017} obtained by exploring a subset of Galileon solutions. Indeed, after the first release of Planck results, all studies were restricted to tracker solutions defined by the relation
\begin{equation}
    \forall a, \quad \Hbar^{2} x = \mathrm{const} = \Hbar_{\infty}^{2} x_{\infty}
\end{equation}
where the subscript $\infty$ stands for the limit $a \rightarrow \infty$. Tracker solutions are particularly interesting because they are attractor scenarios, first exhibited in \cite{DeFelice-2010}, which means that every Galileon scenario will eventually converge towards a universe with the same properties and evolution as a particular tracker solution. Furthermore, the background evolution of tracker solutions can be determined analytically which helps to speed up very much the exploration of the parameter space. Finally, restricting to tracker solutions imposes a new relation on Galileon parameters, which allows one of the free parameters to be removed and reduces the size of the parameter space:
\begin{equation}
    c_{2} - 6c_{3} + 18c_{4} - 15c_{5} - 6c_{G} = 0 \label{eq:tracker}
\end{equation}
All Galileon scenarios considered in our work start their evolution away from a tracker one since they are very unlikely to follow exactly relation \eqref{eq:tracker}. However, they all eventually converge towards a particular tracker evolution when $\Hbar^{2}x$ gets sufficiently close to the associated tracker value $\Hbar_{\infty}^{2} x_{\infty}$. Thus, our approach includes scenarios that reach their associated tracker well before the Galileon has any significant physical impact, which cannot be distinguished from tracker solutions. Since we have flat priors on the $c_{i}$'s, we let cosmological data drive the fit and determine whether best-fit Galileon scenarios reach their associated tracker early or not. \\
\newline
It was argued in \cite{Barreira-2013,Barreira-2014} that the only Galileon scenarios that fit correctly the temperature CMB power spectrum are those that reached their associated tracker solution well before the dark energy dominated era, i.e. before $a \lesssim 0.5$, and that there is no loss of generality by restricting to purely tracker solutions because the Galileon field does not modify significantly the physics of the early Universe. We wanted to check this statement with more recent cosmological data. Furthermore, those tracker only studies concluded to tensions between the Galileon model and BAO data. Allowing for more scenarios with a richer phenomenology, especially for the cosmological background which determines BAO observables, is important to avoid missing possibly interesting sets of parameters. \\
\begin{figure}[htbp]
\begin{center}
\begin{subfigure}{0.5\textwidth}
  \centering
  \includegraphics[width=\linewidth]{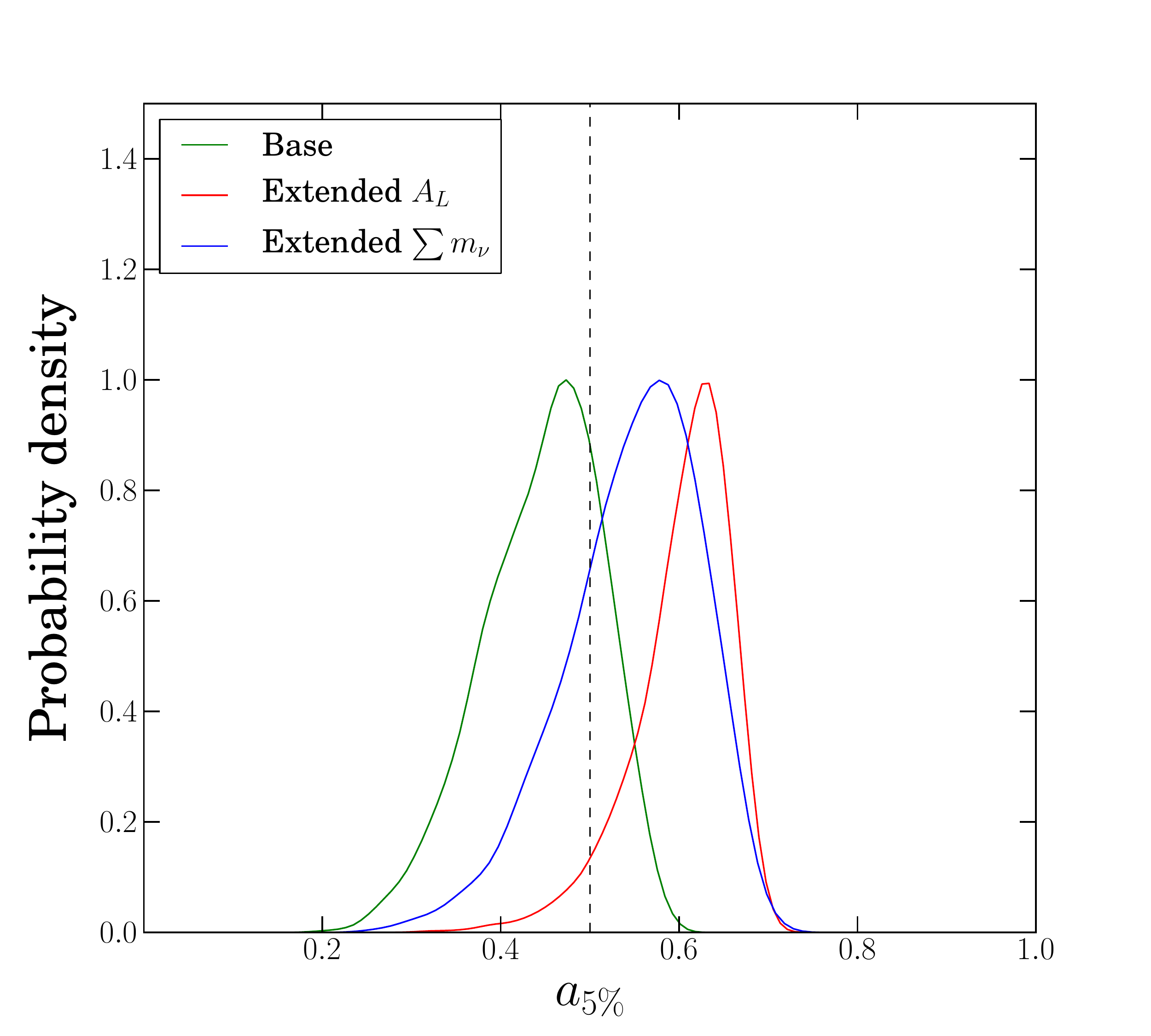}
\end{subfigure}%
\begin{subfigure}{0.5\textwidth}
  \centering
  \includegraphics[width=\linewidth]{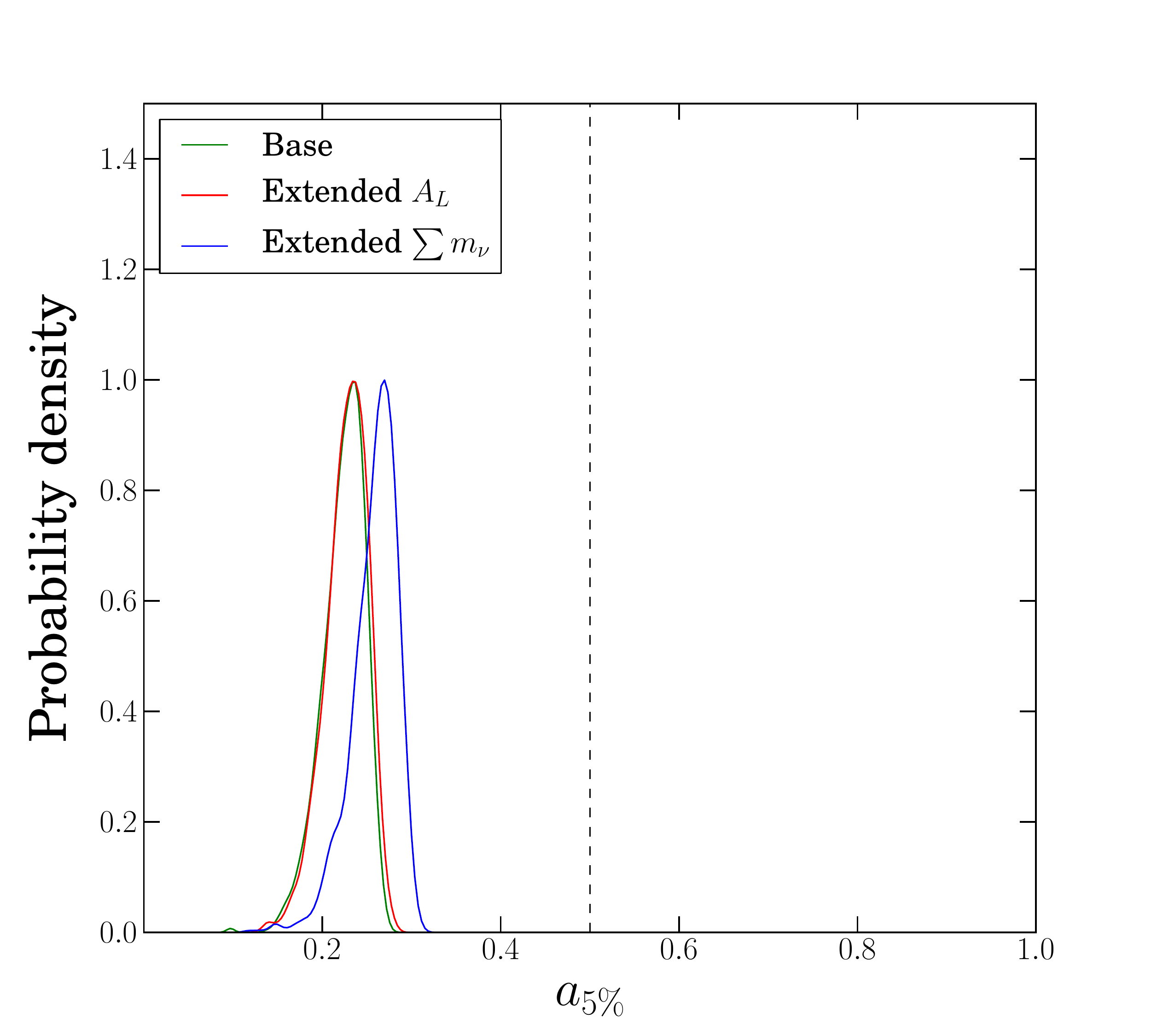}
\end{subfigure}
\end{center}
\caption{Probability density of the derived parameter $a_{5\%}$ of the scenarios selected in the Markov chains for the base and extended Full galileon (left) and Cubic galileon (right) models.}
\label{fig:tscale}
\end{figure}
\newline
In order to quantify the moment at which the associated tracker is reached, we define the scale factor $a_{r}$ at which $\left| \frac{\Hbar^{2} x - \Hbar_{\infty}^{2} x_{\infty}}{\Hbar_{\infty}^{2} x_{\infty}} \right| \leq r$. We chose to consider that the tracker evolution is reached when $r=0.05$. In practice, to determine $a_{5\%}$, the asymptotic value $\Hbar_{\infty}^{2} x_{\infty}$ was computed at $a=10$. At this value, we observed that all scenarios selected in the Markov chains had converged to their associated tracker at an excellent precision. The $a_{5\%}$ parameter is to be compared to $a \sim 0.5$ at which the dark energy era starts. According to \cite{Barreira-2014}, data should prefer only scenarios with $a_{5\%} < 0.5$. In Figure \ref{fig:tscale} we show that the Cubic scenarios selected by the MCMC exploration are in this case while data prefer scenarios of the Full galileon model that are not. 
Thus, the scenarios of the Cubic galileon model selected in the Markov chains of our general analysis behave as suggested in \cite{Barreira-2013,Barreira-2014} and reach their associated tracker solution well before the dark energy dominated era. This is an important result since the Cubic galileon model reduced to tracker solutions was ruled out by \cite{Renk-2017} with a significance of $\sim 8 \sigma$, using the ISW effect, and also by \cite{Peirone-2017} using weak lensing data. \\
\newline
Yet, the above statement does not hold in the case of the Full galileon model, for which the associated tracker is reached later. Most of the selected scenarios reach their associated tracker at the transition to the dark energy era (base models), or after it started (extended models), as shown in Figure \ref{fig:tscale}. In particular, only the best-fit of the base model reaches the tracker solution at better than 5\% at $a \sim 0.5$ (see Figure \ref{fig:trackers}).  Examination of the Markov chains revealed that this tendency is driven by BAO data that favour scenarios that are further away from their associated tracker. 
\newline
Most of these scenarios, and in particular best-fit scenarios, would be missed by a tracker-only study despite the fact that they fit the CMB temperature power spectrum quite correctly, as shown by the CMB contribution to the $\chi^2$ values of the global best-fits in tables \ref{table:Contraintes CMB+BAO+JLA},\ref{table:Contraintes CMB+BAO+JLA_Al},\ref{table:Contraintes CMB+BAO+JLA_mnu}.

\section{Conclusion}
\label{section:section7}

In this paper, we explored and constrained the general solution of the full covariant Galileon model with direct disformal coupling to matter, as well as its uncoupled Cubic version. We used cosmological observations from recent CMB, BAO and SNIa data, and from the observation of GW170817. We found that looking at general solutions of the Galileon model, instead of restricting to tracker solutions only, was important to avoid missing interesting scenarios. Indeed, our best-fit scenarios in the Full galileon model provide good fits to CMB spectra while reaching their associated tracker solution at or after the start of the dark energy dominated era in contrast with the argument in favor of limiting the exploration of the Galileon model to tracker solutions only. Our results show that tracker solutions are not completely representative of the sub-space of solutions that are able to reproduce CMB data. As recent works proved that tracker solutions could not provide good fits to those data \cite{Barreira-2014,Renk-2017,Peirone-2017}, it was essential to explore the general solution to establish robust constraints on the Galileon model. \\
\newline
However, despite providing good agreement with CMB spectra, the general solutions of the Galileon models, either Full or Cubic, are not able to reproduce simultaneously CMB, BAO and JLA data. In addition, the best fitting scenarios have very low values of the optical depth at reionization and thus of the reionization redshift, which are incompatible with direct astrophysical observations of reionization. The preference for low values of the optical depth at reionization was found to be related to the higher lensing effect in Galileon best-fit scenarios compared to $\Lambda$CDM ones. This led us to consider Galileon models extended to one additional parameter that impacts the predicted lensing power spectrum. We studied extensions to either $A_{L}$, the lensing spectrum normalization, or $\sum m_{\nu}$, the sum of active neutrino masses. These extensions remove the incompatibility with reionization data, but the tension with BAO measurements remains, although reduced. Note that whatever the fit (base or extended model) the uncoupled Cubic model always performs worse than the Full galileon model, due to poorer agreement with CMB data. \\
\newline
Finally, constraints from the time delay measured between GW170817 and its electromagnetic counterpart were applied \textit{a posteriori} on all scenarios selected by the Markov chains of the Full galileon model, which predicts a difference between GW and light speeds, contrary to the uncoupled Cubic model. All scenarios were found to be fully excluded by this measurement. 
Thus, altogether, the general solution of the Galileon model appears to be ruled out by recent observations. This work extends and confirms the constraints obtained with tracker solutions only \cite{Barreira-2014,Barreira-2014b,Renk-2017,Peirone-2017}. 

\section{Acknowledgement}
\label{section:section8}

We thank Alexandre Barreira for the interesting discussions about the features of the Galileon model, as well as for kindly helping us reproduce previous results of his team. We also thank Anthony Lewis for providing open access to and help with \CAMB \ and \CosmoMC.

\bibliographystyle{unsrt}

\begin{thebibliography}{99}

\bibitem{Nicolis-2009}
Nicolis A., Rattazzi R., Trincherini E., \emph{The Galileon as a local modification of gravity}, \emph{Phys. Rev.} {\bf D79} (2009), 064036.

\bibitem{Deffayet-2009a}
Deffayet C., Esposito-Farese G., Vikman A., \emph{Covariant Galileon}, \emph{Phys. Rev. D} {\bf 79} (2009a), 084003.

\bibitem{Deffayet-2009b}
Deffayet C., Deser S., Esposito-Farese G., \emph{Generalized Galileons: All scalar models whose curved background extensions maintain second-order field equations and stress tensors}, \emph{Phys. Rev. D} {\bf 80} (2009b), 064015.

\bibitem{Appleby-2011}
Appleby S., Linder E.V., \emph{The Paths of Gravity in Galileon Cosmology}, {\bf 1203} \emph{JCAP} (2012), 043.

\bibitem{Barreira-2013}
Barreira A., Li B., Sanchez A., Baugh C.M., Pascoli S., \emph{Parameter space in Galileon gravity models}, \emph{Phys. Rev.} {\bf D87} (2013), 103511.

\bibitem{Neveu-2013}
Neveu J. et al., \emph{Experimental constraints on the uncoupled Galileon model from SNLS3 data and other cosmological probes}, \emph{A \& A} {\bf 555} (2013), A53.

\bibitem{Neveu-2014}
Neveu J. et al., \emph{First experimental constraints on the disformally coupled Galileon model}, \emph{A \& A} {\bf 569} (2014), A90.

\bibitem{Neveu-2016}
Neveu J. et al., \emph{Constraining the $\Lambda$CDM and Galileon models with recent cosmological data}, \emph{A \& A} {\bf 600} (2017), A40.

\bibitem{Barreira-2014}
Barreira A., Li B., Baugh C., Pascoli S., \emph{The observational status of Galileon gravity after Planck}, \emph{JCAP} {\bf 1408} (2014), 059.

\bibitem{Barreira-2014b}
Barreira A., Li B., Baugh C.M., Pascoli S., \emph{$\nu$Galileon: modified gravity with massive neutrinos as a testable alternative to $\Lambda$CDM}, \emph{Phys. Rev.} {\bf D90} (2014), 023528.

\bibitem{Renk-2017}
Renk J., Zumalac\'arregui M., Montanari F., Barreira A., \emph{Galileon gravity in light of ISW, CMB, BAO and H$_0$ data}, \emph{JCAP} {\bf 1710} (2017), 10.

\bibitem{Peirone-2017}
Peirone M., Frusciante N., Hu B., Raveri M., Silvestri A., \emph{Do current cosmological observations rule out all Covariant Galileons?}, \emph{Phys. Rev.} {\bf D97} (2018), 063518.

\bibitem{Ezquiaga-2017}
Ezquiaga J.M., Zumalac\'arregui M., \emph{Dark Energy After GW170817: Dead Ends and the Road Ahead}, \emph{Phys. Rev. Lett.} {\bf 119} (2017), 25.

\bibitem{Wang-2017}
Wang H. et al., \emph{The GW170817/GRB 170817A/AT 2017gfo Association: Some Implications for Physics and Astrophysics}, \emph{Astrophys. J.} {\bf 851} (2017), L18.

\bibitem{Sakstein-2017}
Sakstein J., Jain B., \emph{Implications of the Neutron Star Merger GW170817 for Cosmological Scalar-Tensor Theories}, \emph{Phys. Rev. Lett.} {\bf 119} (2017), 251303.

\bibitem{Brax-2012}
Brax P., Burrage C., Davis A.-C., \emph{Shining Light on Modifications of Gravity}, \emph{JCAP} {\bf 1210} (2012), 016.

\bibitem{Burrage-2010}
Burrage C., Seery D., \emph{Revisiting fifth forces in the Galileon model}, \emph{JCAP} {\bf 1008} (2010), 011.

\bibitem{Barreira-2012}
Barreira A., Li B., Baugh C.M., Pascoli S., \emph{Linear perturbations in Galileon gravity models}, \emph{Phys. Rev.} {\bf D86} (2012), 124016.

\bibitem{Challinor-1998}
Challinor A., Lasenby A., \emph{Cosmic microwave background anisotropies in the CDM model: A Covariant and gauge invariant approach}, \emph{Astrophys. J.} {\bf 513} (1999), 1-22.

\bibitem{camb_notes}
Lewis A., \emph{CAMB Notes}, \url{https://cosmologist.info/notes/CAMB.pdf}.

\bibitem{Ma-1994}
Ma C.-P., Bertschinger E., \emph{Cosmological perturbation theory in the synchronous versus conformal Newtonian gauge}, \emph{Astrophys. J.} {\bf 455} (1995), 7-25.

\bibitem{Betoule-2014}
Betoule M. et al., \emph{Improved cosmological constraints from a joint analysis of the SDSS-II and SNLS supernova samples}, \emph{A \& A} {\bf 568} (2014), A22.

\bibitem{Beutler-2011}
Beutler F. et al., \emph{The 6dF Galaxy Survey: Baryon Acoustic Oscillations and the Local Hubble Constant}, \emph{Mon. Not. Roy. Astron. Soc.} {\bf 416} (2011).

\bibitem{Ross-2014}
Ross A.J. et al., \emph{The clustering of the SDSS DR7 main Galaxy sample – I. A 4 per cent distance measure at $z = 0.15$}, \emph{Mon. Not. Roy. Astron. Soc.} {\bf 449} (2015), 1.

\bibitem{Alam-2016}
BOSS Collaboration, \emph{The clustering of galaxies in the completed SDSS-III Baryon Oscillation Spectroscopic Survey: cosmological analysis of the DR12 galaxy sample}, \emph{Mon. Not. Roy. Astron. Soc.} {\bf 470} (2017), 3.

\bibitem{Planck-2015}
Planck Collaboration, \emph{Planck 2015 results. XIII. Cosmological parameters}, \emph{A \& A} {\bf 594} (2016), A13.



\bibitem{GBM-2017}
Abbott B.P. et al., \emph{Multi-messenger Observations of a Binary Neutron Star Merger}, \emph{Astrophys. J.} {\bf 848} (2017), 2.

\bibitem{DeFelice-2011}
De Felice A., Tsujikawa S., \emph{Generalized Galileon cosmology}, \emph{Phys. Rev.} {\bf D84} (2011), 124029.

\bibitem{DeFelice-2010}
De Felice A., Tsujikawa S., \emph{Cosmology of a covariant Galileon field}, \emph{Phys. Rev. Lett.} {\bf 105} (2010), 111301.

\bibitem{Lewis-2002}
Lewis A., Bridle S., \emph{Cosmological parameters from CMB and other data: A Monte Carlo approach}, \emph{Physical Review D} {\bf 66} (2002), 103511.

\bibitem{Bouwens-2015}
Bouwens R.J. et al., \emph{Reionization after Planck: The Derived Growth of the Cosmic Ionizing Emissivity now matches the Growth of the Galaxy UV Luminosity Density}, \emph{Astrophys. J.} {\bf 811} (2015), 2.

\bibitem{Ade-2013}
Planck Collaboration, \emph{Planck 2013 results. XVII. Gravitational lensing by large-scale structure}, \emph{A \& A} {\bf 571} (2014), A17.

\bibitem{Ade-2015}
Planck Collaboration, \emph{Planck 2015 results. XV. Gravitational lensing}, \emph{A \& A} {\bf 594} (2016), A15. 

\bibitem{Planck-2018}
Planck Collaboration, \emph{Planck 2018 results. VI. Cosmological parameters}, [arXiv:1807.06209], submitted to \emph{A \& A}. 

\bibitem{Kajita-1998}
Kajita T., \emph{Atmospheric neutrino results from Super-Kamiokande and Kamiokande: Evidence for neutrino(mu) oscillations}, \emph{Proceedings, International Conference on Neutrino physics and astrophysics (Neutrino'98): Takayama, Japan, June 4-9, 1998} {\bf 77} (1999), 123-132.

\bibitem{Kraus-2004}
Kraus C. et al., \emph{Final results from phase II of the Mainz neutrino mass search in tritium beta decay}, \emph{Eur. Phys. J.} {\bf C40} (2005), 447-468.

\bibitem{Aseev-2011}
Aseev V.N. et al., \emph{An upper limit on electron antineutrino mass from Troitsk experiment}, \emph{Phys. Rev.} {\bf D84} (2011), 112003.

\bibitem{Barreira-2015}
Barreira A., Brax P., Clesse S., Li B., Valageas P., \emph{Linear perturbations in K-mouflage cosmologies with massive neutrinos}, \emph{Phys. Rev.} {\bf D91} (2015), 063528.

\bibitem{Bellomo-2017}
Bellomo N., Bellini E., Hu B., Jimenez R., Pena-Garay C., Verde L., \emph{Hiding neutrino mass in modified gravity cosmologies}, \emph{JCAP} {\bf 1702} (2017), 043.

\bibitem{Baldi-2013}
Baldi M., Villaescusa-Navarro F., Viel M., Puchwein E.,
Springel V., Moscardini L., \emph{Cosmic Degeneracies I: Joint N-body Simulations of Modified Gravity and Massive Neutrinos}, \emph{Mon. Not. Roy. Astronm. Soc.} {\bf 440} (2014), 75.

\bibitem{Dirian-2017}
Dirian Y., \emph{Changing the Bayesian prior: Absolute neutrino mass constraints in nonlocal gravity}, \emph{Phys. Rev.} {\bf D96} (2017), 083513.

\bibitem{Riess-2016}
Riess A. et al., \emph{A 2.4\% Determination of the Local Value of the Hubble Constant}, \emph{Astrophys. J.} {\bf 826} (2016), 1.

\bibitem{Planck-2016}
Planck collaboration, \emph{Planck 2015 results. XIV. Dark energy and modified gravity}, \emph{A \& A} {\bf 594} (2016), A14.  


\end{thebibliography}

\appendix
\section{Galileon perturbation equations}
\label{appendix:appendixA}

We detail here the scalar perturbation equations for the Galileon quantities and the equation of motion of the Galileon field scalar perturbation, from \cite{Barreira-2012}, rewritten with our notations:
\begin{enumerate}
    \item $\chi^{G} = \fchi_{1} \cdot \gamma + \fchi_{2} \cdot \gamma' + \frac{1}{\kappa a^{2}} \left( \fchi_{3} \cdot k\Hcal\Zcal + \fchi_{4} \cdot k^{2}\eta \right)$ with :
\begin{eqnarray}
	f_{1}^{\chi} & = & \frac{k^{2}}{\kappa a^{2}} \left[ -2\frac{c_{3}}{a^{2}}x^{2}\Hbarcal^{2} + 12\frac{c_{4}}{a^{4}}x^{3} \Hbarcal^{4} - 15\frac{c_{5}}{a^{6}}x^{4}\Hbarcal^{6} - 4\frac{c_{G}}{a^{2}}x\Hbarcal^{2} \right] \\
	f_{2}^{\chi} & = & \frac{H_{0}}{\kappa a^{2}} \left[ c_{2}x\Hbarcal - 18\frac{c_{3}}{a^{2}}x^{2}\Hbarcal^{3} + 90\frac{c_{4}}{a^{4}}x^{3}\Hbarcal^{5} - 105\frac{c_{5}}{a^{6}}x^{4}\Hbarcal^{7} - 18\frac{c_{G}}{a^{2}}x\Hbarcal^{3} \right] \\
	f_{3}^{\chi} & = & -2\frac{c_{3}}{a^{2}}x^{3}\Hbarcal^{2} + 15\frac{c_{4}}{a^{4}}x^{4}\Hbarcal^{4} - 21\frac{c_{5}}{a^{6}}x^{5}\Hbarcal^{6} - 6\frac{c_{G}}{a^{2}}x^{2}\Hbarcal^{2} \\
	f_{4}^{\chi} & = & \frac{3}{2}\frac{c_{4}}{a^{4}}x^{4}\Hbarcal^{4} - 3\frac{c_{5}}{a^{6}}x5\Hbarcal^{6} - \frac{c_{G}}{a^{2}}x^{2}\Hbarcal^{2}
\end{eqnarray}

    \item $q^{G} = \fq_{1} + \frac{1}{\kappa a^{2}} \fq_{2} \cdot k^{2} \left( \sigma - \Zcal \right)$ with :
\begin{eqnarray}
	f_{1}^{q} & = & \frac{k}{\kappa a^{2}} \left[ c_{2} H_{0}x\Hbarcal\gabar - \frac{c_{3}}{a^{2}} \left( -2x^{2}\Hbar^{2}\gabar' + 6H_{0}x^{2}\Hbarcal^{3}\gabar \right) + \frac{c_{4}}{a^{4}} \left( -12x^{3}\Hbarcal^{4}\gabar' + 18H_{0}x^{3}\Hbarcal^{5}\gabar \right) \right. \nonumber \\
	&& \left. - \frac{c_{5}}{a^{6}} \left( -15x^{4}\Hbarcal^{6}\gabar' + 15H_{0}x^{4}\Hbarcal^{7}\gabar \right) - \frac{c_{G}}{a^{2}} \left( -4x\Hbarcal^{2}\gabar' + 6H_{0}x\Hbarcal^{3}\gabar \right) \right] \\
	f_{2}^{q} & = & \frac{c_{4}}{a^{4}}x^{4}\Hbarcal^{4} - 2\frac{c_{5}}{a^{6}}x^{5}\Hbarcal^{6} - \frac{2}{3}\frac{c_{G}}{a^{2}}x^{2}\Hbarcal^{2}
\end{eqnarray}

    \item $\Pi^{G} = \fPi_{1} + \frac{1}{\kappa a^{2}} \left( \fPi_{2} \cdot k\Hcal\sigma - \fPi_{3} \cdot k\sigma' + \fPi_{4} \cdot k^{2}\phi \right)$ with :
\begin{eqnarray}
	f_{1}^{\Pi} & = & \frac{k^{2}}{\kappa a^{2}} \left[ \frac{c_{4}}{a^{4}} \left( 4x^{3}\Hbarcal^{4}\gabar - 6x^{2}\Hbarcal^{3}\cir{\left( x\Hbarcal \right)}\gabar \right) - \frac{c_{5}}{a^{6}} \left( 12x^{4}\Hbarcal^{6}\gabar - 3x^{4}\Hbarcal^{5}\cir{\Hbarcal}\gabar - 12x^{3}\Hbarcal^{5}\cir{\left( x\Hbarcal \right)}\gabar \right) \right. \nonumber \\
	&& \left. + 2\frac{c_{G}}{a^{2}}\Hbar\cir{\left( x\Hbarcal \right)}\gabar \right] \\
	f_{2}^{\Pi} & = & \frac{c_{4}}{a^{4}} \left( 3x^{4}\Hbarcal^{4} - 6x^{3}\Hbarcal^{3}\cir{\left( x\Hbarcal \right)} \right) - \frac{c_{5}}{a^{6}} \left( 12x^{5}\Hbarcal^{6} - 3x^{5}\Hbarcal^{5}\cir{\Hbarcal} - 15x^{4}\Hbarcal^{5}\cir{\left( x\Hbarcal \right)} \right) \nonumber \\
	&& + 2\frac{c_{G}}{a^{2}}x\Hbar\cir{\left( x\Hbarcal \right)} \\
	f_{3}^{\Pi} & = & \frac{c_{4}}{a^{4}}x^{4}\Hbarcal^{4} + 3\frac{c_{5}}{a^{6}}x^{4}\Hbar^{5}\cir{\left( x\Hbarcal \right)} \\
	f_{4}^{\Pi} & = & -\frac{c_{4}}{a^{4}}x^{4}\Hbarcal^{4} - \frac{c_{5}}{a^{6}} \left( -6x^{5}\Hbarcal^{6} + 6x^{4}\Hbar^{5}\cir{\left( x\Hbarcal \right)} \right) + 2\frac{c_{G}}{a^{2}}x^{2}\Hbarcal^{2}
\end{eqnarray}

    \item $0 = \feom_{1} \cdot \gabar'' + \feom_{2} \cdot \gabar' + \feom_{3} \cdot k^{2}\gabar + \feom_{4} \cdot k\Hcal\Zcal + \feom_{5} \cdot k\Zcal' + \feom_{6} \cdot k^{2}\eta$ with :
\begin{eqnarray}
	f_{1}^{eom} & = & c_{2} - 12\frac{c_{3}}{a^{2}}x\Hbarcal^{2} + 54\frac{c_{4}}{a^{4}}x^{2}\Hbarcal^{4} - 60\frac{c_{5}}{a^{6}}x^{3}\Hbarcal^{6} - 6\frac{c_{G}}{a^{2}}\Hbarcal^{2} \\
	f_{2}^{eom} & = & H_{0} \left[ 2c_{2}\Hbarcal - \frac{c_{3}}{a^{2}} \left( 12x\Hbarcal^{2}\cir{\Hbarcal} + 12\Hbarcal^{2}\cir{\left( x\Hbarcal \right)} \right) + \frac{c_{4}}{a^{4}} \left( -108x^{2}\Hbarcal^{5} + 108x^{2}\Hbarcal^{4}\cir{\Hbarcal} + 108x\Hbarcal^{4}\cir{\left( x\Hbarcal \right)} \right) \right. \nonumber \\
	&& \left. - \frac{c_{5}}{a^{6}} \left( -240x^{3}\Hbarcal^{7} + 180x^{3}\Hbarcal^{6}\cir{\Hbarcal} + 180x^{2}\Hbarcal^{6}\cir{\left( x\Hbarcal \right)} \right) - 12\frac{c_{G}}{a^{2}}\Hbarcal^{2}\cir{\Hbarcal}  \right] \\
	f_{3}^{eom} & = & c_{2} - \frac{c_{3}}{a^{2}} \left( 4x\Hbarcal^{2} + 4\Hbarcal\cir{\left( x\Hbarcal \right)} \right) + \frac{c_{4}}{a^{4}} \left( -10x^{2}\Hbarcal^{4} + 12x^{2}\Hbarcal^{3}\cir{\Hbarcal} + 24x\Hbarcal^{3}\cir{\left( x\Hbarcal \right)} \right) \nonumber \\
	&& - \frac{c_{5}}{a^{6}} \left( -36x^{3}\Hbarcal^{6} + 24x^{3}\Hbarcal^{5}\cir{\left( \Hbarcal \right)} + 36x^{2}\Hbarcal^{5}\cir{\left( x\Hbarcal \right)} \right) - \frac{c_{G}}{a^{2}} \left( 2\Hbarcal \right) \\
	f_{4}^{eom} & = & c_{2}x - \frac{c_{3}}{a^{2}} \left( 6x^{2}\Hbarcal^{2} + 4x\Hbarcal\cir{\left( x\Hbarcal \right)} \right) + \frac{c_{4}}{a^{4}} \left( -6x^{3}\Hbarcal^{4} + 12x^{3}\Hbarcal^{3}\cir{\Hbarcal} + 36x^{2}\Hbarcal^{3}\cir{\left( x\Hbarcal \right)} \right) \nonumber \\
	&& - \frac{c_{5}}{a^{6}} \left( -45x^{4}\Hbarcal^{6} + 30x^{4}\Hbarcal^{5}\cir{\Hbarcal} + 60x^{3}\Hbarcal^{5}\cir{\left( x\Hbarcal \right)} \right) - \frac{c_{G}}{a^{2}} \left( 6x\Hbarcal^{2} + 4x\Hbarcal\cir{\Hbarcal} + 4\Hbarcal\cir{\left( x\Hbarcal \right)} \right) \\
	f_{5}^{eom} & = & -2\frac{c_{3}}{a^{2}}x^{2}\Hbarcal^{2} + 12\frac{c_{4}}{a^{4}}x^{2}\Hbarcal^{4} - 15\frac{c_{5}}{a^{6}}x^{4}\Hbarcal^{6} - 4\frac{c_{G}}{a^{2}}x\Hbarcal^{2} \\
	f_{6}^{eom} & = & \frac{c_{4}}{a^{4}} \left( -4x^{3}\Hbarcal^{4} + 6x^{2}\Hbarcal^{3}\cir{\left( x\Hbarcal \right)} \right) - \frac{c_{5}}{a^{6}} \left( -12x^{4}\Hbarcal^{6} + 3x^{4}\Hbarcal^{5}\cir{\Hbarcal} + 12x^{3}\Hbarcal^{5}\cir{\left( x\Hbarcal \right)} \right) \nonumber \\
	&& - 2\frac{c_{G}}{a^{2}}\Hbarcal\cir{\left( x\Hbarcal \right)}
\end{eqnarray}

\end{enumerate}

\section{MCMC parameter space exploration}
\label{appendix:appendixB}

We started the parameter space exploration from the best-fit published in \cite{Neveu-2016} (resp. \cite{Barreira-2014}) for the Full (resp. Cubic) galileon model with priors from the Planck Collaboration for the set of parameters common to the three models. This first step provided more relevant priors (best-fit and covariance matrix) for a second run. For each Galileon model and combination of data, we ran eight chains in parallel, with approximately 10,000 points per chain from which we got rid of the first half to remove the burn-in phase. We ran only four chains for the $\Lambda$CDM model, to save computation time, and we checked that our results were compatible with those from the Planck Collaboration \cite{Planck-2015}. For all models, the convergence of the chains was checked using the Gelman-Rubin diagnostic $R$, and was assumed to be reached when $R-1 < 0.03$. \\
\newline
We were careful to remove non-viable scenarios during the parameter space exploration. Indeed, some sets of parameters introduce ghost degrees of freedom, or Laplace instabilities. The theoretical conditions required to avoid these particular sets were derived in \cite{DeFelice-2011,Neveu-2016}, for both scalar and tensorial perturbations. Note that the theoretical conditions use only background quantities. So, computing them is very fast and it is possible to check for the good behaviour of tensorial perturbations even though their evolution and power spectrum are not computed explicitly.
In practice, our modified version of \CosmoMC \ first checks that the theoretical constraints are fullfilled when hopping to a new point. If they are not, the point is immediately rejected before computing the perturbations evolution. \\
\newline
Finally, the constraints from the combination of CMB, BAO and JLA data were obtained by adding constraints from the JLA sample using the method of Importance Sampling to chains that used CMB and BAO data simultaneously. The constraints from GW170817 were obtained by computing \textit{a posteriori} the time delay predicted by each scenario saved in the chains ran on cosmological data.

\end{document}